\documentclass[prb,twocolumn,showpacs,amsmath,amssymb]{revtex4}
\usepackage{amssymb}
\usepackage{txfonts}
\usepackage{bbm}
\usepackage{graphicx}
\usepackage{appendix}
\usepackage{epsf}
\usepackage{epstopdf}
\usepackage{amsmath}
\usepackage[usenames]{color}

\begin{document}

\title{Impurity-induced quasiparticle interference in the parent compounds of iron-pnictide superconductors}

\author
{Huaixiang Huang,$^{1,2}$  Yi Gao$^{1,3}$, Degang Zhang,$^{1}$ and
C. S. Ting$^{1,4}$}

\affiliation{$^{1}$Texas Center for Superconductivity and Department of
Physics, University of Houston, Houston, Texas 77204, USA\\
$^{2}$Department of Physics, Shanghai University , Shanghai 200444, China\\
$^{3}$Department of Physics and Institute of Theoretical Physics, Nanjing Normal University, Nanjing, Jiangsu 210046, China\\
$^{4}$Department of Physics, Fudan University, Shanghai 200433, China\\
}

\date{\today}
\begin{abstract}
The impurity-induced quasiparticle interference(QPI) in the parent
compounds of iron-pnictide superconductors is investigated based on
a phenomenological two-orbital four-band model and T-matrix method.
We find the QPI is sensitive to the value of the magnetic order
which may vary from one compound to another. For small value of the
magnetic order, the pattern of oscillation in the local density of
states (LDOS) induced by the QPI exhibits two-dimensional
characteristics, consistent with the standing wave state observed in
the $1111$ compound. For larger value of the magnetic order, the
main feature of the spatial modulation of the LDOS is the existence
of one-dimensional stripe structure which is in agreement with the
nematic structure in the parent compound of the $122$ system. In
both cases the system shows $C_2$ symmetry and only in the larger
magnetic order case, there exist in-gap bound states. The
corresponding QPI in $q$-space is also presented. The patterns of
modulation in the LDOS at nonzero energies are attributed to the
interplay between the underlying band structure and Fermi surfaces.

\end{abstract}
 \maketitle

\narrowtext
\section{introduction}
One of the intriguing issues in condensed matter physics is the
recent discovery of the new families of iron-based high-T$_{c}$
superconductors.~\cite{1,2,3,4,5,one1} Like cuprates,
superconductivity arises from electron- or hole- doping of their
antiferromagnetic(AFM) parent compounds. Different from cuprates
whose parent compounds are Mott insulators, the parent compounds of
the iron-pnictides are bad metals whose resistivity is several
orders of magnitude larger than that of normal metals. When lowing
the temperature, accompanied by the formation of
spin-density-wave(SDW) state,~\cite{two1,two2,sdw,jun} the parent
compounds of $122$ ($\mathrm{AFe_2As_2}$) undergo a tetragonal to
orthorhombic structural transition while for $1111$
($\mathrm{RFeAsO_xF_y}$) the temperature for the structural
transition is higher than that for the AFM transition.
Neutron-Diffraction measurements~\cite{neudiff1} show that the
magnetic moment ($0.87\mu_B$) per Fe at $5\mathrm{K}$ in
$\mathrm{BaFe_2As_2}$ is substantially larger than the moment
($0.36\mu_B$) per Fe in $\mathrm{LaFeAsO}$.~\cite{4}

Experiments on $1111$ compound~\cite{helmut,changliu,lxyang} show an
extra larger hole pocket around the $\Gamma$ point, different from
other pnictides due to surface polarity or surface-driven electronic
structure. It is believed that the SDW picture established in the
$122$ compound could apply to the $1111$ compound as well. The Fermi
surfaces(FSs) of the iron-pnictides have disconnected sheets, the
strong nesting between the hole FSs around the $\Gamma$ point and
the electron ones around the $\mathrm{M}$ point induces the SDW
instability. The formation of SDW will partially gap the FSs and
will have great effect on the QPI as well as other physical
properties. Before unveiling the origin of superconductivity,
understanding the ordered parent compound is an important task.

Scanning tunneling microscopy (STM) on pnictides has provided much
useful information on the electronic
properties.~\cite{stm1,stm2,stm3} For underdoped $122$ compound
$\mathrm{Ca(Fe_{1-x}Co)_2As_2}$, a remarkable nematic electronic
structure~\cite{chuang} has been observed by means of
spectroscopic-imaging (SI) STM, where a one-dimensional (1D)
structure aligns along the crystal a-axis with antiparallel spins,
and the QPI imaging disperses predominantly along the b-axis of the
material. While for the parent compound of $1111$ system
$\mathrm{LaOFeAs}$, there are two types of surface after cleavage.
STM studies have revealed that a two-dimensional (2D) strong
standing wave pattern~\cite{xiaodong} induced by the QPI appears in
one type of surface. And the corresponding dispersions along $q_x$
and $q_y$ are similar. Diversified electronic structure states have
been observed in various systems,~\cite{na0,th} whether they play a
key role in the mechanism of superconductivity~\cite{na1,na2} has
triggered much attention and motivated our investigation.

Previously, based on some effective models, the QPI has been
calculated by using different
methods.~\cite{han,jk,eug,aak,mazin,zho2} In order to well explain
the above mentioned experiments, we solve exactly the QPI patterns
as well as their Fourier component based on a phenomenological
model~\cite{zhang} which considers the asymmetry of the As atoms
above and below the Fe-Fe plane. Due to the surface effect during
cleavage, the heights of the As atoms above and below the Fe-Fe
plane may not be equal to each other, thus this model is suitable to
study the properties of the surface layers in the iron-pnictides and
should be more appropriate to describe the STM experiments which are
surface-sensitive. Our study will be within the framework of
T-matrix method adopted by previous
works.~\cite{zhang,zjx,zhang1,zhang2} The most interesting result is
that we can obtain 1D and 2D modulations of the LDOS by employing
the same model. The value of the magnetic order has great effect on
the spatial modulation of the LDOS. Small value of magnetic order
leads to 2D pattern while larger magnetic order results in 1D
structure. In both cases, the patterns exhibit $C_2$ symmetry. To
the best of our knowledge, there still lack works concerning both of
the interference patterns.

The paper is organized as follows. In Sec. II, we introduce the
model and work out the formalism.  In Sec. III, we show the property
of the normal state in the presence of impurity. In Sec. IV, we
study the modulation of the LDOS induced by QPI with small value of
magnetic order. In Sec. V, we investigate the QPI in the case of
larger value of magnetic order. Finally, we give a summary .

\section{model and formalism}

\begin{figure}
\includegraphics[width=9cm]{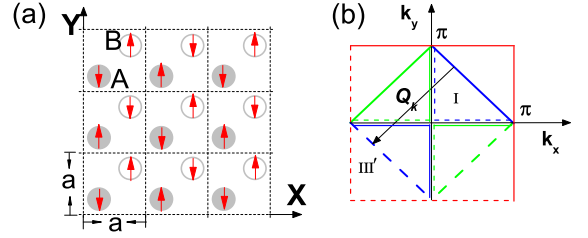}
\caption{(color online) (a) Schematic lattice structure of the Fe
layer in the SDW state. A and B are the two inequivalent Fe ions.
(b) The first Brillouin zone (red line) and the magnetic Brillouin
zone (inner square). The I and III quadrants are embraced by the
blue lines while the II and IV quadrants are embraced by the green
ones. All the dashed lines are not included in those quadrants. The
corresponding $Q_k$ of moving $k$ within $\mathrm{I}$ to that within
$\mathrm{III}^{\prime}$ is $(-\pi,-\pi)$. In the area inside the
blue (green) lines the corresponding $V_k=1$ $(-1)$. } \label{fig1}
\end{figure}

We start with the two-orbital four-band tight-binding
phenomenological model~\cite{zhang} which takes the $d_{xz}$ and
$d_{yz}$ orbitals of Fe ions into account and each unit cell
accommodates two inequivalent Fe ions. The reason we adopt this
model is that such a minimal model reproduces qualitatively the
evolution of the FSs observed by angle resolved photoemission
spectroscopy (ARPES) experiments~\cite{14,15,16,17,19} in both the
electron- and hole- doped iron-pnictides. Based on this model, the
obtained phase diagram~\cite{zho}, dynamic spin
susceptibility~\cite{gao1}, Andreev bound state inside the vortex
core~\cite{gao2} as well as domain wall structure~\cite{huang} are
all consistent with the ARPES~\cite{arpes}, neutron
scattering~\cite{ns} and STM~\cite{stm} experiments. The
tight-binding part of the Hamiltonian can be expressed as
\begin{eqnarray}
H_{0}=&-\sum_{i\tilde{\nu} j\nu\sigma}(t_{i\tilde{\nu}
j\nu}c^\dagger_{i\tilde{\nu}\sigma}c_{j\nu\sigma}+h.c.)-t_0\sum_{{\bf
i}\nu\sigma}c^{\dagger}_{i\nu\sigma}c_{i\nu\sigma} ,
\end{eqnarray}
where $i,j$ are the site indices, $\tilde{\nu},\nu=0,1$ are the
orbital indices, and $t_0$ is the chemical potential. $t_1$
represents the nearest-neighbor (nn) hopping between the same
orbitals on Fe ions, $t_2$ and $t_3$ denote next-nearest-neighbor
(nnn) hoppings between the same orbitals mediated by the up and down
As ions, respectively. $t_4$ is the nnn hopping between different
orbitals.  In this paper, we adopted the hopping parameters as in
Ref.~\onlinecite{zhang}, i.e. $t_1=0.5ev$, $t_2=0.4t_1$,
$t_3=-2.0t_1$, and $t_4=0.04t_1$. In the following the energy and
scattering potential are measured in unit of $t_1$. The distance
between the nnn Fe ions is $a$ and set as unit, which is shown in
Fig.~\ref{fig1}(a).

With the formation of SDW, the first Brillouin zone (BZ) needs to be
folded into the magnetic Brillouin zone (MBZ). The Fourier
transformation of the electron destruction operator can be written
as
\begin{eqnarray}\label{2}
c_{\mathfrak{P}i\nu\sigma}&=&\frac{1}{\sqrt{N}}\sum_{k}[c_{\mathfrak{P}\nu
k\sigma }e^{i\textbf{k}\cdot R_{\bf{i}}}+ c_{\mathfrak{P}\nu
k+Q_k\sigma }e^{i(\textbf{k}+Q_k)\cdot
R_{\bf{i}}}]\nonumber\\
&=&\frac{1}{\sqrt{N}}\sum_{k}[c^{0}_{\mathfrak{P}\nu k\sigma
}e^{i\textbf{k}\cdot R_{\bf{i}}}+ c^{1}_{\mathfrak{P}\nu k\sigma
}e^{i(\textbf{k}+Q_k)\cdot R_{\bf{i}}}],
\end{eqnarray}
where $N$ is the number of unit cells, $\mathfrak{P}$ can be
sublattice A or B corresponding to the two inequivalent Fe ions, $k$
is restricted in the MBZ and $Q_k$ is chosen to be $\pm(\pi,\pi)$ or
$\pm(\pi,-\pi)$, depending on which quadrant $k$ belongs to, such
that $k+Q_k$ is in the first BZ. We have shown an example in
Fig.~\ref{fig1}(b). Here $c^{0}_{\mathfrak{P}\nu k\sigma}$
represents $c_{\mathfrak{P}\nu\sigma k}$, and
$c^{1}_{\mathfrak{P}\nu k\sigma}$ represents $c_{\mathfrak{P}\nu
k+Q_k \sigma}$. Define $C^{\dag\alpha }_{\sigma}(k)=(c^{\dag
\alpha}_{A0k\sigma},c^{\dag \alpha
}_{A1k\sigma},c^{\dag\alpha}_{B0k\sigma},c^{\dag \alpha }_{B1k\sigma
})$, then $H_0=\sum_{k\sigma \alpha}C^{\dag\alpha }_{\sigma}
M_k^{\alpha} C^{\alpha}_{\sigma}$, in which $\alpha=0,1$ and
\begin{eqnarray}\label{3}
M_k^{0}&=&\left(
  \begin{array}{cccc}
    a_1 & a_3 & a_4 & \,0 \\
    a_3 & a_1 & \,0 & a_4 \\
    a_4 & \,0 & a_2 & a_3 \\
    \,0 & a_4 & a_3 & a_2 \\
  \end{array}
\right),
\end{eqnarray}
 where $a_1=-2t_2\cos{k_y}-2t_3\cos{k_x}$,
 $ a_2=-2t_3\cos{k_y}-2t_2\cos{k_x}$,
 $ a_3=-2t_4(\cos{k_x}+\cos{k_y})$,
 $a_4=-2t_1(\cos{\frac{k_x+k_y}{2}}+\cos{\frac{k_y-k_x}{2}})$. For $M_k^1$ the corresponding $k$ changes into $k+Q_k$. After diagonalizing
the above Hamiltonian, we obtain $H_0= \sum_{\mu\nu\alpha
k}\epsilon^{\alpha}_{\mu\nu}(k)\psi^{\dag\alpha}_{\mu\nu}(k)\psi^{\alpha}_{\mu\nu}(k)$,
with the energy band indices $\alpha,\mu,\nu$ being $0$ or $1$. The
analytical expressions for the eight energy bands can be written as
$\epsilon^0_{\mu\nu}(k)=\frac{1}{2}(a_1+a_2)+(-1)^{\nu}a_3+(-1)^{\mu}\Gamma-t_0$,
$\epsilon^1_{\mu\nu}(k)=\epsilon^0_{\mu\nu}(k+Q_k)$ with
$\Gamma=\sqrt{\Gamma_1^2+a_4^2}$, $\Gamma_1=\frac{1}{2}(a_1-a_2)$,
and they do not depend on the spin. At half filling $t_0=-0.622$ and
the canonical transformation matrix reads
\begin{equation}\label{4}
\left(
  \begin{array}{c}
    c^{\alpha}_{A0k\sigma }\\
    c^{\alpha}_{A1k\sigma }\\
    c^{\alpha}_{B0k\sigma }\\
    c^{\alpha}_{B1k\sigma }\\
  \end{array}
\right)=\left(
          \begin{array}{cccc}
            b^{\alpha}_{00}  &-b^{\alpha}_{01}  &b^{\alpha}_{10}  &-b^{\alpha}_{11} \\
            b^{\alpha}_{00}   &\;b^{\alpha}_{01}  &b^{\alpha}_{10}  & \;b^{\alpha}_{11} \\
            b^{\alpha}_{11}   &\;b^{\alpha}_{10}  &b^{\alpha}_{01}  &\; b^{\alpha}_{00} \\
            b^{\alpha}_{11}  &-b^{\alpha}_{10}  &b^{\alpha}_{01}  &-b^{\alpha}_{00} \\
          \end{array}
        \right)
        \left(
          \begin{array}{c}
            \psi^{\alpha}_{00\sigma}(k) \\
            \psi^{\alpha}_{01\sigma}(k)\\
            \psi^{\alpha}_{10\sigma}(k)\\
            \psi^{\alpha}_{11\sigma}(k)\\
          \end{array}
        \right).
\end{equation}
The matrix elements are functions of $k$,
$b^0_{00}=x_1(\Gamma+\Gamma_1)$, $ b^0_{01}=x_2a_4$,
$b^0_{10}=x_2(\Gamma_1-\Gamma)$, $ b^0_{11}=x_1a_4$ and
$x_{1,2}=[2(\Gamma_1\pm\Gamma)^2+2a_4^2]^{-1/2}$ are the
renormalization factors. Substituting Eq. (\ref{4}) into Eq.
(\ref{2}), we have
\begin{eqnarray}\label{bogoliutran}
\label{5}c^{\alpha}_{A\nu k\sigma
}&=&\sum_{\mu^{\prime},\nu^{\prime}}(-1)^{\nu^{\prime}(1-\nu)}
b^{\alpha}_{\mu^{\prime}\nu^{\prime}}(k)
\psi^{\alpha}_{\mu^{\prime}\nu^{\prime}\sigma}(k),\\
\label{6}c^{\alpha}_{B\nu k\sigma
}&=&\sum_{\mu^{\prime},\nu^{\prime}}(-1)^{\nu^{\prime}\nu}b^{\alpha}_{\mu^{\prime}+1\nu^{\prime}+1}(k)
\psi^{\alpha}_{\mu^{\prime}\nu^{\prime}\sigma}(k).
\end{eqnarray}
In our coordinates, the origin is located at the A sublattice, the
configuration of magnetic order is shown in Fig.~\ref{fig1}(a). Note
that magnetic order
$\tilde{m}\sum_{\mu}(n_{i\mu\uparrow}-n_{i\mu\downarrow})$
varies~\cite{zho} with site, $\tilde{m}=me^{iQ_k\cdot R_{A}}$ for
sublattice $A$ and $me^{iQ_k\cdot(R_{B}+\hat{x}/2-\hat{y}/2)}$ for
sublattice $B$. The experimentally observed SDW~\cite{zho} term is
introduced as,
\begin{eqnarray}\label{hsdw}
H_{SDW}&=&m\;\;\sum_{\nu\alpha k \sigma}  \sigma[c^{\dag
\alpha}_{A\nu k\sigma} c^{\alpha+1}_{A\nu k\sigma}+V_k c^{\dag
\alpha}_{B\nu k\sigma} c^{\alpha+1}_{B\nu k\sigma}],\nonumber\\
&=&2m\!\sum_{\alpha\nu\mu\mu^{\prime}k\sigma} \sigma \
B^{\alpha\mu\nu } _{\mu^{\prime}}(k)
\psi^{\dag\alpha}_{\mu\nu\sigma}(k)\psi^{\alpha+1}_{\mu^{\prime}\nu\sigma}(k)\\\label{7}
B^{\alpha\mu\nu}  _{\mu^{\prime}}&=&b^{\alpha}_{\mu\nu}
b^{\alpha+1}_{\mu^{\prime}\nu}+V_k b^{\alpha}_{\mu+1\nu+1}
b^{\alpha+1}_{\mu^{\prime}+1\nu+1}, \label{8}
\end{eqnarray}
$V_k=1$ for the first and third quadrants, otherwise $V_k=-1$, as
shown in Fig.1(b). $\sigma$ being $\pm1$ corresponds to spin up and
spin down, respectively. A single impurity $V_s
\sum_{\nu\sigma}c^{\dag
}_{A0\nu\sigma}c_{A0\nu\sigma}+V_m\sum_{\nu\sigma}\sigma c^{\dag
}_{A0\nu\sigma}c_{A0\nu\sigma}$ is located at the origin in
sublattice A, the Hamiltonian of the impurity part can be written as
\begin{eqnarray}\label{himp}
H_{imp}=\frac{2}{N} \sum_{^{\mu\alpha\nu\sigma k}
_{\mu^{\prime}\alpha^{\prime} k^{\prime}}}(V_s+\sigma V_m)
b^{\alpha}_{\mu\nu}(k)b^{\alpha^{\prime}}_{\mu^{\prime}\nu}(k^{\prime})
\psi^{\dag\alpha}_{\mu\nu\sigma}(k)\psi^{\alpha^{\prime}}_{\mu^{\prime}\nu\sigma}(
k^{\prime}),
\end{eqnarray}
where $V_s$ and $V_m$ represent the nonmagnetic part and magnetic
part of the impurity potential, respectively. The total Hamiltonian
is $H=H_0+H_{SDW}+H_{imp}$. In the following we will solve the QPI
state.

 Define the two-point Green's function as
\begin{eqnarray}\label{10}
G^{\alpha\mu\nu\sigma}_{\alpha^{\prime}\mu^{\prime}\nu^{\prime}\sigma^{\prime}}(k,k^{\prime};i
\omega_n)=-\mathcal{F}\langle
\mathrm{T}_{\tau}\psi^{\alpha}_{\mu\nu\sigma
}(k,\tau)\psi^{\dag\alpha^{\prime}}_{\mu^{\prime}\nu^{\prime}\sigma^{\prime}
}(k^{\prime},0) \rangle,
\end{eqnarray}
where $\mathcal{F}\phi(\tau)$ denotes the Fourier transform of
$\phi(\tau)$ in Matsubara frequencies, $\mathrm{T}_{\tau}$ is the
time ordering operator and $\psi(\tau)=e^{\tau H}\psi e^{-\tau H}$.
By using the equation of motion for Green's function and
$\frac{\partial \psi(\tau)}{\partial \tau}=e^{\tau H}[H,\psi]
e^{-\tau H}$  we obtain
\begin{eqnarray}
\label{11}G^{\alpha\mu\nu\sigma}_{\alpha^{\prime}\mu^{\prime}\nu^{\prime}\sigma^{\prime}}
   (k,k^{\prime};i\omega_n)&=&mG^{0}_{\alpha\mu\nu} \sigma\!\!\! \sum
   _{\mu^{\prime\prime}}B^{\alpha\mu\nu
   }_{\mu^{\prime\prime}}(k)G^{\alpha+1\mu^{\prime\prime}\nu\sigma}_{\alpha^{\prime}\mu^{\prime}\nu^{\prime}\sigma^{\prime}}
   (k,k^{\prime};i
   \omega_n)\nonumber\\
   &+&g^{\alpha\mu\nu\sigma}_{\alpha^{\prime}\mu^{\prime}\nu^{\prime}\sigma^{\prime}}
   (k,k^{\prime};i\omega_n),\\
\label{12}g^{\alpha\mu\nu\sigma}_{\alpha^{\prime}\mu^{\prime}\nu^{\prime}\sigma^{\prime}}
   (k,k^{\prime};i\omega_n)&=&G^{0}_{\alpha\mu\nu}(k,i\omega_n)\delta_{\alpha\alpha^{\prime}}
   \delta_{\mu\mu^{\prime}}\delta_{\nu\nu^{\prime}}\delta_{\sigma\sigma^{\prime}}\delta_{kk^{\prime}}+\nonumber\\
   \frac{2}{N}(V_s\!\!+\!\sigma\!
   V_m)\!\!\!&G&\!\!\!^{0}_{\alpha\mu\nu}(k,i\omega_n)b^{\alpha}_{\mu\nu}(k)
   D^{\nu\sigma}_{\alpha^{\prime}\mu^{\prime}\nu^{\prime}\sigma^{\prime}}(k^{\prime},i\omega_n),\\
\label{13}D^{\nu\sigma}_{\alpha^{\prime}\mu^{\prime}\nu^{\prime}\sigma^{\prime}}(k^{\prime},i\omega)&=&\!\!\!\sum_{\alpha^{\prime\prime}\mu^{\prime\prime}
k^{\prime\prime}}\!\! b^{\alpha^{\prime\prime}}_{
\mu^{\prime\prime}\nu} (k^{\prime\prime})
G^{\alpha^{\prime\prime}\mu^{\prime\prime}\nu\sigma}_{\alpha^{\prime}\mu^{\prime}\nu^{\prime}\sigma^{\prime}}(k^{\prime\prime},
k^{\prime};i \omega_n),
\end{eqnarray}
where $G^0_{\alpha\mu\nu}(k,i \omega_n)=[i
\omega_n-\epsilon^{\alpha}_{\mu\nu}(k)]^{-1}$ is the bare Green's
function. Since the translational invariance is broken by the
impurity, the Green's function depends on two momenta $k$ and
$k^{\prime}$. Solving the Green's function is the basis for
calculating the LDOS, to this end, we introduce a $4\times4$ matrix
$\mathbb{S}$,
\begin{eqnarray}\label{smatr}
  \begin{array}{c}
    \mathbb{S}  \\
  \end{array}
 \left(
  \begin{array}{c}
    G^{0 0 \nu \sigma}_{\mu^{\prime}\nu^{\prime}\alpha^{\prime}\sigma^{\prime}} \\
    G^{0 1 \nu \sigma}_{\mu^{\prime}\nu^{\prime}\alpha^{\prime}\sigma^{\prime}} \\
    G^{1 0\nu \sigma}_{\mu^{\prime}\nu^{\prime}\alpha^{\prime}\sigma^{\prime}} \\
    G^{1 1\nu \sigma}_{\mu^{\prime}\nu^{\prime}\alpha^{\prime}\sigma^{\prime}} \\
  \end{array}
\right)=\left(
          \begin{array}{c}
            g^{0 0 \nu \sigma}_{\mu^{\prime}\nu^{\prime}\alpha^{\prime}\sigma^{\prime}} \\
            g^{0 1 \nu \sigma}_{\mu^{\prime}\nu^{\prime}\alpha^{\prime}\sigma^{\prime}} \\
            g^{1 0\nu \sigma}_{\mu^{\prime}\nu^{\prime}\alpha^{\prime}\sigma^{\prime}} \\
            g^{1 1 \nu \sigma}_{\mu^{\prime}\nu^{\prime}\alpha^{\prime}\sigma^{\prime}} \\
          \end{array}
        \right),\nonumber\\
   \mathbb{S}=\left(
    \begin{array}{cc}
      I & -m\sigma S_0\\
     -m\sigma S_1 & I\\
    \end{array}
  \right),
 \end{eqnarray}
where $I$ is the $2\times2$ unit matrix and
\begin{eqnarray}\label{smatralpha}
 S_{\alpha}=  \left(
      \begin{array}{cc}
        B^{\alpha0\nu}_{0}G^0_{\alpha0\nu} &B^{\alpha0\nu}_{1} G^0_{\alpha0\nu} \\
        B^{\alpha1\nu}_{0}G^0_{\alpha1\nu} &B^{\alpha1\nu}_{1}G^0_{\alpha1\nu}  \\
      \end{array}
    \right).
\end{eqnarray}
From Eqs. (\ref{smatr}) and (\ref{smatralpha}), we finally obtain
the Green's function
\begin{eqnarray}
\label{16}G^{\alpha\mu\nu\sigma}_{\alpha^{\prime}\mu^{\prime}\nu^{\prime}\sigma^{\prime}}(k,k^{\prime}\!\!\!&;&\!\!\!i\omega_n)
         =\Lambda^{\alpha\mu}_{\alpha^{\prime}\mu^{\prime}}(\nu\sigma k)G^0_{\alpha^{\prime}\mu^{\prime}\nu}(k,i
              \omega_n)\delta_{\nu\nu^{\prime}}\delta_{\sigma\sigma^{\prime}}\delta_{kk^{\prime}}\nonumber\\
         &+&\!\frac{2}{N}(V_s\!+\!\sigma V_m)D^{\nu\sigma}_{\alpha^{\prime}\mu^{\prime}\nu^{\prime}\sigma^{\prime}}(k^{\prime},i
        \omega_n)f_1^{\alpha\mu\nu\sigma}(k),\\
\label{17}f_1^{\alpha\mu\nu\sigma}(k)\!&=&\!\!\sum_{\alpha^{\prime\prime}\mu^{\prime\prime}}\Lambda^{\alpha\mu}_{\alpha^{\prime\prime}\mu^{\prime\prime}}(\nu\sigma
k)b^{\alpha^{\prime\prime}}_{\mu^{\prime\prime}\nu}(k)G^{0}_{\alpha^{\prime\prime}\mu^{\prime\prime}\nu}(k,i\omega_n),
\end{eqnarray}
where $\Lambda=\mathbb{S}^{-1}$ is a matrix, the upper index
$\alpha\mu$ denote the row of the elements and the lower ones denote
the column of the elements. At this stage
$D^{\nu\sigma}_{\alpha^{\prime}\mu^{\prime}\nu^{\prime}\sigma^{\prime}}(k^{\prime},i
        \omega_n)$ is still unknown. Combining Eqs. (\ref{13})
and (\ref{16}), we obtain
\begin{eqnarray}\label{18}
D^{\nu\sigma}_{\alpha^{\prime}\mu^{\prime}}(k^{\prime},i
  \omega_n)&=&f_{4}^{\alpha^{\prime}\mu^{\prime}\nu^{\prime}\sigma^{\prime}}(k^{\prime})[1-2(V_s\!\!+\!\!\sigma
   V_m)f_2^{\nu\sigma}]^{-1},
\end{eqnarray}
in which
\begin{eqnarray}
\label{19}f_2^{\nu\sigma}&=&\frac{1}{N}\sum_{\alpha\mu k}
   b^{\alpha}_{\mu\nu}(k)f_1^{\alpha\mu\nu\sigma}(k),\\
\label{20}f_3^{\alpha^{\prime}\mu^{\prime}\nu\sigma}(k^{\prime})&=&\sum_{\alpha\mu
   }\Lambda^{\alpha\mu}_{\alpha^{\prime}\mu^{\prime}}(\nu\sigma
   k^{\prime})b^{\alpha}_{\mu\nu}(k^{\prime}),\\
\label{21}f_{4}^{\alpha^{\prime}\mu^{\prime}\nu^{\prime}\sigma^{\prime}}(k^{\prime})
  &=&f_3^{\alpha^{\prime}\mu^{\prime}\nu\sigma}(k^{\prime})G^0_{\alpha^{\prime}\mu^{\prime}\nu}(k^{\prime},i
  \omega_n)\delta_{\nu\nu^{\prime}}\delta_{\sigma\sigma^{\prime}}.
\end{eqnarray}
 The poles of the Green's function consist
of the poles of the bare $G^0$ and poles of
$D^{\nu\sigma}_{\alpha^{\prime}\mu^{\prime}}(k^{\prime},
  i \omega_n)$, the latter ones signify the appearance
of new bound states  due to the impurity. From
Eqs.(\ref{18})(\ref{21}) we can see that the index
$\nu^{\prime}\sigma^{\prime}$ of $D$ can be omitted. We also note
the pole of $D^{\nu\sigma}_{\alpha^{\prime}\mu^{\prime}}(k^{\prime},
i\omega_n)$ is related to the magnitude of the magnetic order since
it contains $\Lambda$. If we consider only the diagonal term of
$\Lambda$, then
$G^{\alpha\mu\nu\sigma}_{\alpha^{\prime}\mu^{\prime}\nu^{\prime}\sigma^{\prime}}(k,k^{\prime};i
\omega_n)=G^0_{\alpha\mu\nu}(k,i
              \omega_n)\delta_{kk^{\prime}}+G^0_{\alpha\mu\nu}(k,i
              \omega_n)T_{matr}G^0_{\alpha^{\prime}\mu^{\prime}\nu}(k^{\prime},i
              \omega_n)$, which is in the form of Dyson's equation. In real space,
the LDOS of each site is
$\rho(r_i,\omega)=-\frac{1}{\pi}\Sigma_{\nu\sigma}\mathrm{Im}[-\mathcal{F}\langle
T_{\tau} c_{\mathfrak{P}i\nu\sigma}(\tau)c^{\dag
}_{\mathfrak{P}i\nu\sigma}(0)\rangle]$. Note
\begin{eqnarray}
\label{22}c_{Ai\nu\sigma}(\tau)c^{\dag
}_{Ai\nu\sigma}(0)\!\!\!&=&\!\!\!\sum_{^{\alpha\nu
k\sigma}_{\alpha^{\prime}k^{\prime}}}c^{\alpha}_{A\nu\sigma
  k}(\tau)c^{\dag \alpha^{\prime}}_{A\nu\sigma
  k^{\prime}}(0)\chi^A_{\alpha}\chi^A_{\alpha^{\prime}}e^{ir_A\cdot(k-k^{\prime})},\\
\label{23}c_{Bi\nu\sigma}(\tau)c^{\dag
  }_{Bi\nu\sigma}(0)\!\!\!&=&\!\!\!\sum_{^{\alpha\nu
k\sigma}_{\alpha^{\prime}k^{\prime}}}c^{\alpha}_{B\nu\sigma
  k}(\tau)c^{\dag \alpha^{\prime}}_{B\nu\sigma
  k^{\prime}}(0)\chi^A_{\alpha}\chi^A_{\alpha^{\prime}}\\
  &&\chi_{\alpha k}\chi_{\alpha^{\prime}
 k^{\prime}}e^{ir_B\cdot(k-k^{\prime})},\nonumber
\end{eqnarray}
 we derive the LDOS on
sublattice $A$ and $B$ in real space, respectively. Throughout the
paper we have $\chi^A_{\alpha}=e^{i\alpha r_A\cdot \tilde{q}}$,
$\chi_{\alpha k}=e^{i\alpha Q_k\cdot r_{AB}}$, $r_A=(n_1,n_2)$,
$\tilde{q}=(\pi,\pi)$, $r_{AB}=(0.5,0.5)$, and $n_1,n_2$ are
integers, i.e. coordinates of sites of sublattice A. The LDOS is
obtained via analytic continuation $i\omega_n\rightarrow
\omega+i\eta$ with $\eta$ being a tiny positive number and related
to the lifetime of the quasiparticle. After some calculations, LDOS
in real space can be expressed as follows
\begin{eqnarray}\label{ldos}
\label{24}\rho(r_A,\omega)\!\!\!&=&\!\!\!-\frac{2}{N\pi}\mathrm{Im}
   \sum_{^{\mu\alpha\nu\sigma k}_{\mu^{\prime}\alpha^{\prime}k^{\prime}}}
  [ \Lambda^{\alpha\mu}_{\alpha^{\prime}\mu^{\prime}}G^0_{\alpha^{\prime}\mu^{\prime}\nu}(k,i
   \omega_n)
   b^{\alpha}_{\mu\nu}(k)b^{\alpha^{\prime}}_{\mu^{\prime}\nu}(k)\chi^A_{\alpha}\chi^A_{\alpha^{\prime}}\nonumber\\
&+& \frac{2}{N}(V_s+\sigma V_m)f_1^{\alpha\mu\nu\sigma}(k) b^{\alpha}_{\mu\nu}(k)\chi^A_{\alpha}\nonumber\\
   &D&\!\!\!^{\nu\sigma}_{\alpha^{\prime}\mu^{\prime}}(k^{\prime},i\omega_n)b^{\alpha^{\prime}}_{\mu^{\prime}\nu}(k^{\prime})
   \chi^A_{\alpha^{\prime}}e^{ir_A\cdot(k-k^{\prime})}]|_{i\omega_n\rightarrow \omega+i0^{+}}, \\
\label{25}\rho(r_B,\omega)\!\!\!&=&\!\!\!-\frac{2}{N\pi}\mathrm{Im}
   \sum_{^{\mu\alpha\nu\sigma k}_{\mu^{\prime}\alpha^{\prime}k^{\prime}}}
   [\Lambda^{\alpha\mu}_{\alpha^{\prime}\mu^{\prime}}G^0_{\alpha^{\prime}\mu^{\prime}\nu}(k,i
   \omega_n)\nonumber\\
&\,&
   b^{\alpha}_{\mu+1\nu+1}(k)b^{\alpha^{\prime}}_{\mu^{\prime}+1\nu+1}(k)
   \chi^A_{\alpha}\chi^A_{\alpha^{\prime}}\chi_{\alpha k}\chi_{\alpha^{\prime} k}\nonumber\\
&+&  \frac{2}{N}(V_s+\sigma V_m)f_1^{\alpha\mu\nu\sigma}(k)
b^{\alpha}_{\mu+1\nu+1}(k)
   \chi^A_{\alpha}
   \chi_{\alpha k} \nonumber\\
   D^{\nu\sigma}_{\alpha^{\prime}\mu^{\prime}}(\!\!\!\!&k&\!\!\!\!^{\prime},
   i\omega_n)b^{\alpha^{\prime}}_{\mu^{\prime}+1\nu+1}(k^{\prime})\chi^A_{\alpha^{\prime}}
   \chi_{\alpha^{\prime}k^{\prime}}
   e^{ir_B\cdot(k-k^{\prime})}]|_{i\omega_n\rightarrow \omega+i0^{+}} .
\end{eqnarray}

We can see that the interband scattering only exists in the same
$\nu$ channel. Elastic scattering of quasiparticle mixes states
having same energy but different momenta. The interference between
the incoming and outgoing waves with momenta $k$ and $k^{\prime}$
can give rise to modulation of the LDOS at the wave vector
$q=k-k^{\prime}$. Such kind of interference pattern can be observed
in SI-STM nowadays. The Fourier component of the LDOS (FC-LDOS) can
be written as,
\begin{eqnarray}
\label{26}\rho_q(\omega)&=&\frac{1}{2N}\sum_{r_A r_B}
[\rho(r_A,\omega)e^{i
                   q\cdot r_A}+\rho(r_B,\omega)e^{i q\cdot r_B}]\\
&=&-\frac{1}{2N\pi
   }\sum_{^{\alpha \nu k\sigma}_{\alpha^{\prime} k^{\prime}}}
          ( \mathrm{Im}\{\tilde{g}_1(k)\}
    (\delta^{q}_{0}\delta^{\alpha}_{\alpha^{\prime}}+\delta^{q}_{\tilde{q}}\delta^{\alpha}_{\alpha^{\prime}+1})\nonumber\\
&+&\mathrm{Im}\{\tilde{g}_2(kk^{\prime})\}
  [\delta_{\alpha^{\prime}}^{\alpha} ( \delta_{k^{\prime}}^{k+q}          +\delta_{k^{\prime}}^{k-q})
  +\delta_{\alpha^{\prime}}^{\alpha+1}(\delta_{k^{\prime}}^{k+q+\tilde{q}}+\delta_{k^{\prime}}^{k-q+\tilde{q}})]\nonumber\\
&-&i\mathrm{Re}\{\tilde{g}_2(kk^{\prime})\}
   [\delta_{\alpha^{\prime}}^{\alpha}(\delta_{k^{\prime}}^{k+q}-\delta_{k^{\prime}}^{k-q})
    +\delta_{\alpha^{\prime}}^{\alpha+1}(\delta_{k^{\prime}}^{k+q+\tilde{q}}-\delta_{k^{\prime}}^{k-q+\tilde{q}})]),\nonumber
\end{eqnarray}
where
$\tilde{g}_1(k)=g^{\alpha\alpha^{\prime}}_{a1}(k)+g^{\alpha\alpha^{\prime}}_{b1}(k)$,
$\tilde{g}_2(kk^{\prime})=g^{\nu\sigma\alpha}_{a2}(k)g^{\nu\sigma\alpha^{\prime}}_{a3}(k^{\prime})+g^{\nu\sigma\alpha}_{b2}(k)g^{\nu\sigma\alpha^{\prime}}_{b3}(k^{\prime})$
and
\begin{eqnarray}
\label{27}g^{\alpha\alpha^{\prime}}_{a1}(k)&=&\sum_{\nu\mu\sigma\mu^{\prime}}\Lambda_{\alpha\mu,\alpha^{\prime}\mu^{\prime}}G^0_{\alpha^{\prime}\mu^{\prime}\nu}(k,i
\omega_n)b^{\alpha}_{\mu\nu}(k)b^{\alpha^{\prime}}_{\mu^{\prime}\nu}(k),\\
\label{28}g^{\alpha\alpha^{\prime}}_{b1}(k)&=&\sum_{\nu\mu\sigma\mu^{\prime}}\Lambda^{\alpha\mu}_{\alpha^{\prime}\mu^{\prime}}G^0_{\alpha^{\prime}\mu^{\prime}\nu}(k,i
\omega_n)b^{\alpha}_{\mu\nu}(k)b^{\alpha^{\prime}}_{\mu^{\prime}+1\nu+1}(k),\\
\label{29}g^{\nu\sigma\alpha}_{a2}(k)&=&\frac{2}{N}(V_s+\sigma
V_m)\sum_{\mu}f_1^{\alpha\mu\nu\sigma}(k)b^{\alpha}_{\mu\nu}(k), \\
\label{30}g^{\nu\sigma\alpha}_{b2}(k)&=&\frac{2}{N}(V_s+\sigma
V_m)\sum_{\mu}f_1^{\alpha\mu\nu\sigma}(k)b^{\alpha}_{\mu+1\nu+1}(k) \chi_{\alpha k},\\
\label{31}g^{\nu\sigma\alpha^{\prime}}_{a3}(k^{\prime})&=&\sum_{\mu^{\prime}}D^{\nu\sigma}_{\alpha^{\prime}\mu^{\prime}}(k^{\prime},i\omega_n)
b^{\alpha^{\prime}}_{\mu^{\prime}\nu}(k^{\prime}),\\
\label{32}g^{\nu\sigma\alpha^{\prime}}_{b3}(k^{\prime})&=&\sum_{\mu^{\prime}}D^{\nu\sigma}_{\alpha^{\prime}\mu^{\prime}}(k^{\prime},i\omega_n)
b^{\alpha^{\prime}}_{\mu^{\prime}+1\nu+1}(k^{\prime})\chi_{\alpha^{\prime}
k^{\prime}}.
\end{eqnarray}
Since the QPI in our model has $C_2$ symmetry, which will be seen
clearly in the remaining paper, the last line of Eq.(\ref{26})
 will be zero, thus we only show the
absolute value of the real part in the corresponding figures of
FC-LDOS. The map is confined in the first BZ and we perform our
calculation with $N=800\times800$ unit cells. We neglect the
component of $\tilde{g}_1(k)$ since we want to see the QPI induced
by impurity clearly. Different from the superconducting phase in
which magnetic impurity and nonmagnetic one have distinct effect on
the LDOS.~\cite{zhang} In the SDW state our calculations show that
the effect of a pure magnetic impurity ($\sigma V_m$) is very
similar to that of a pure nonmagnetic one ($V_s$). While for the
mixed scattering potential $V_s+\sigma V_m$, the effect of the
magnetic part is similar to varying the value of magnetic order, so
in the following, we consider only the QPI induced by nonmagnetic
impurity with different values of magnetic order.


\begin{figure}
\centering
      \includegraphics[width=1.2in]{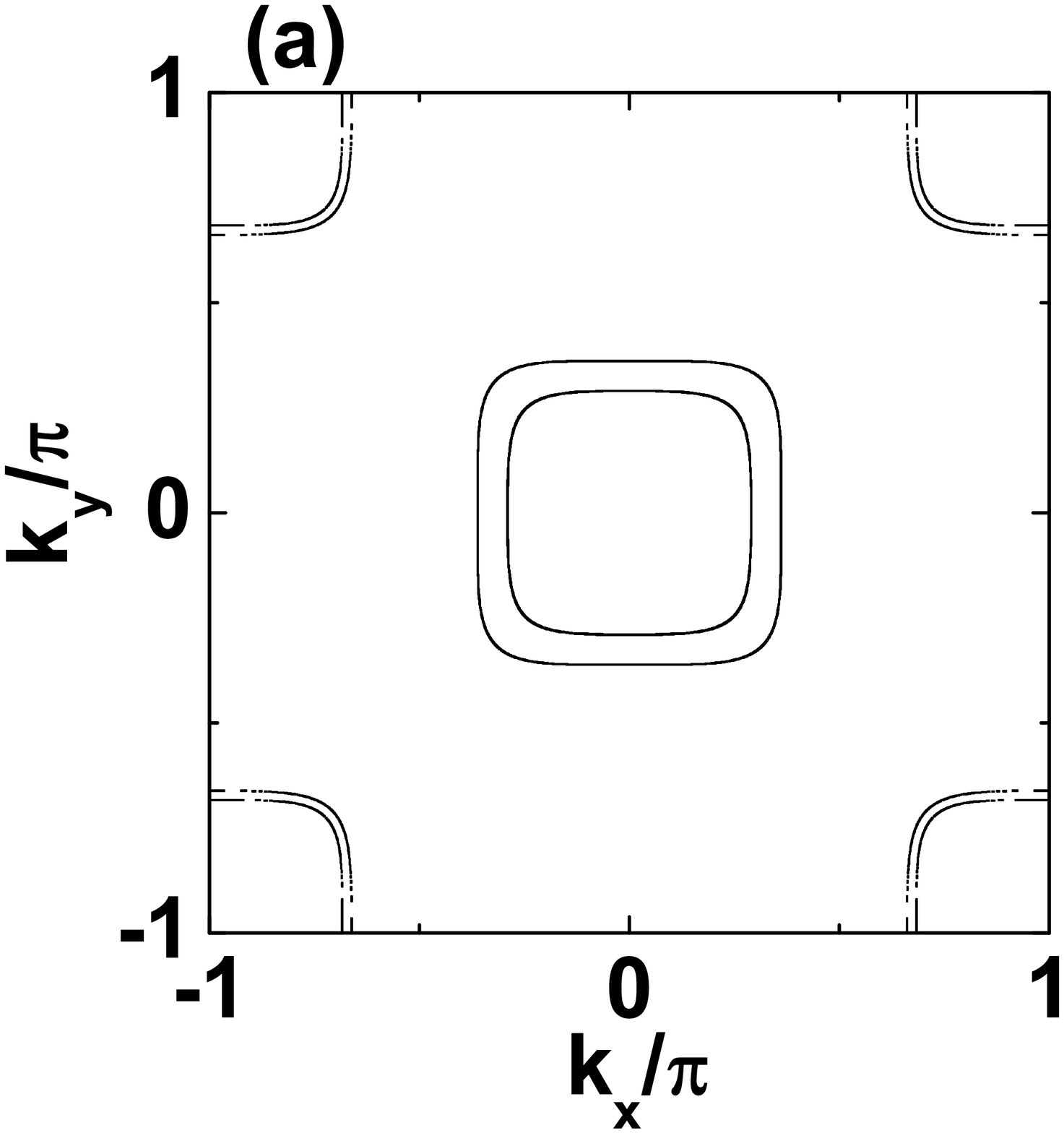}
      \includegraphics[width=1in]{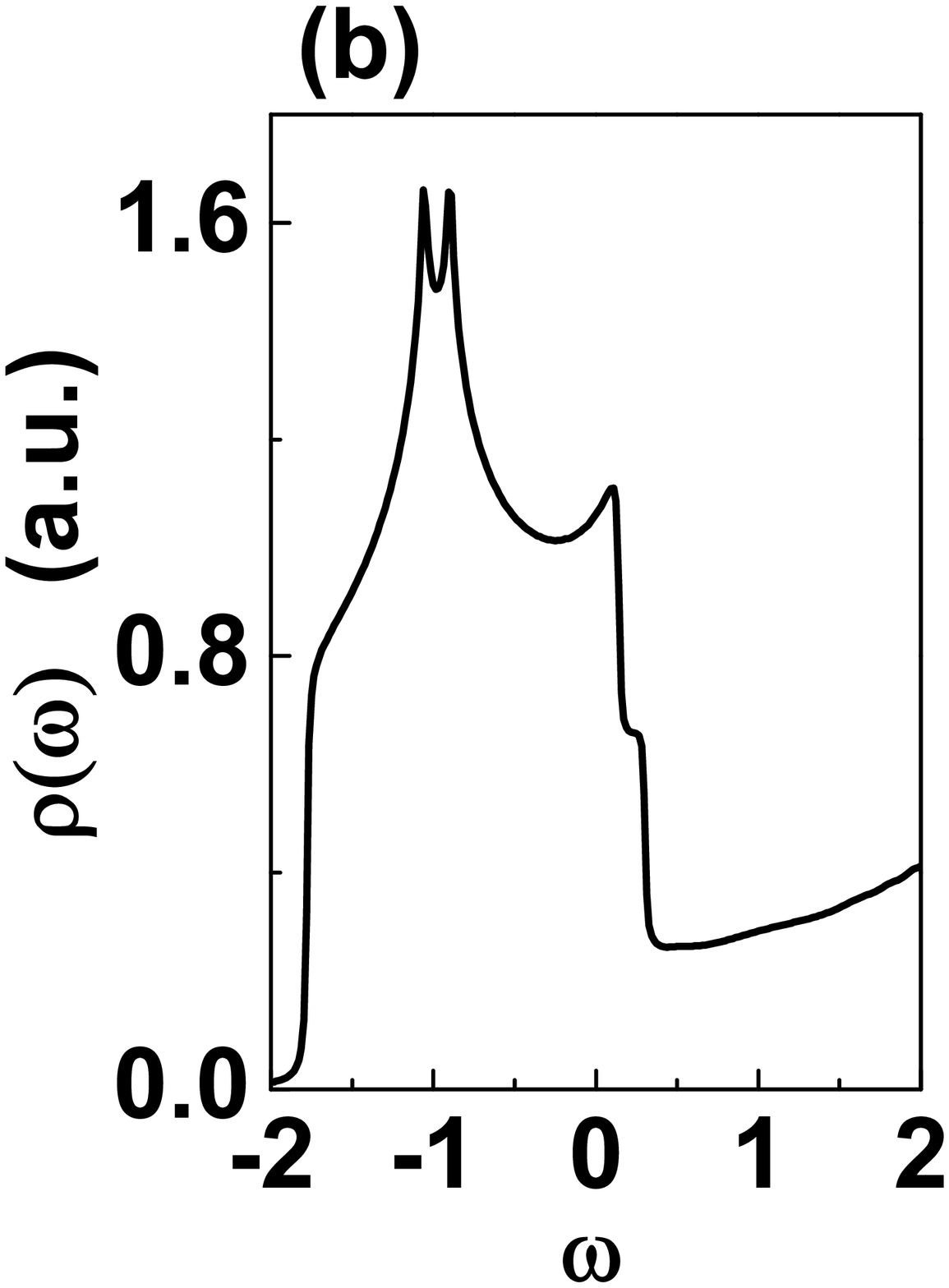}
      \includegraphics[width=1in]{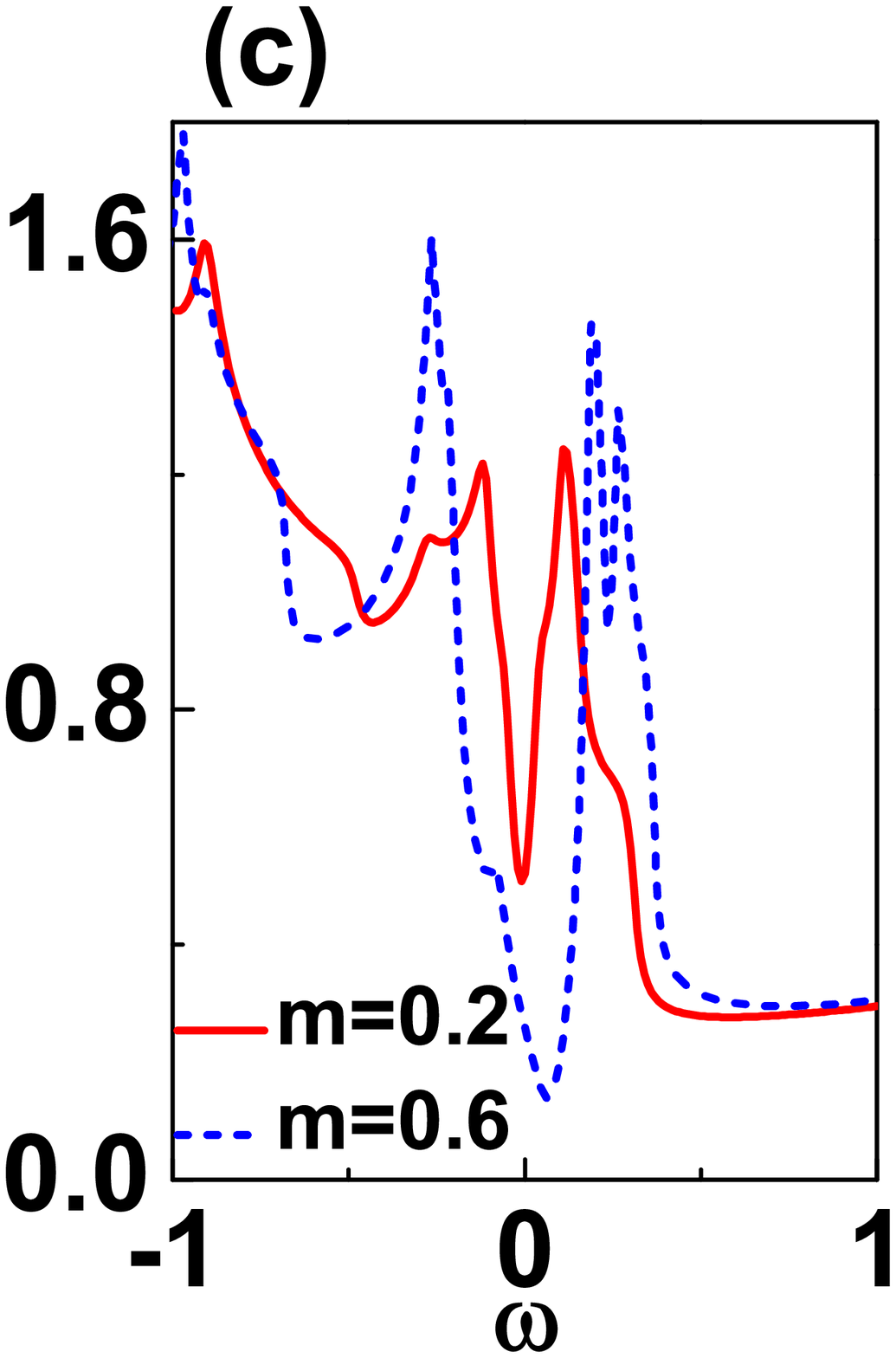}
      \includegraphics[width=1.65in]{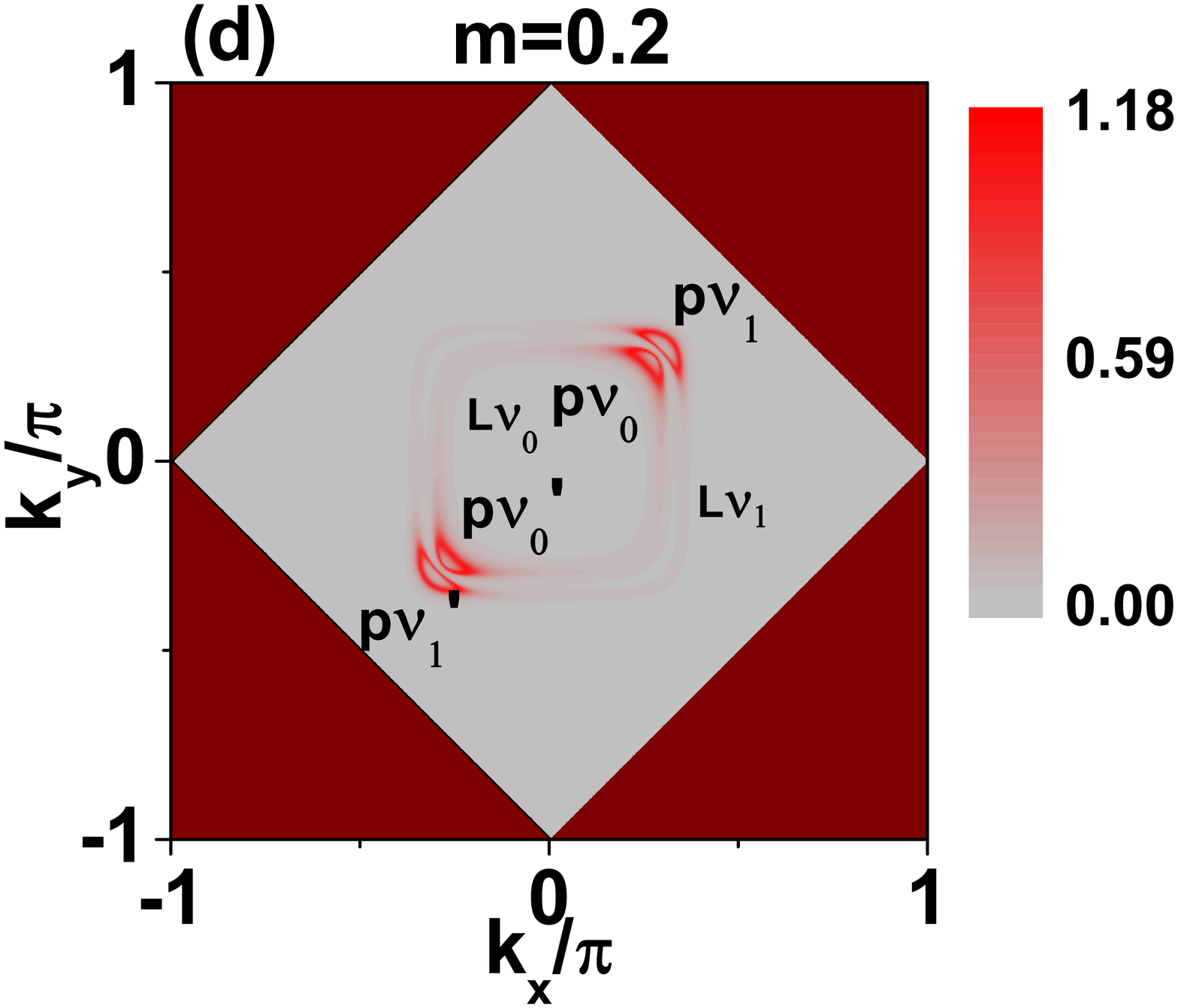}
      \includegraphics[width=1.65in]{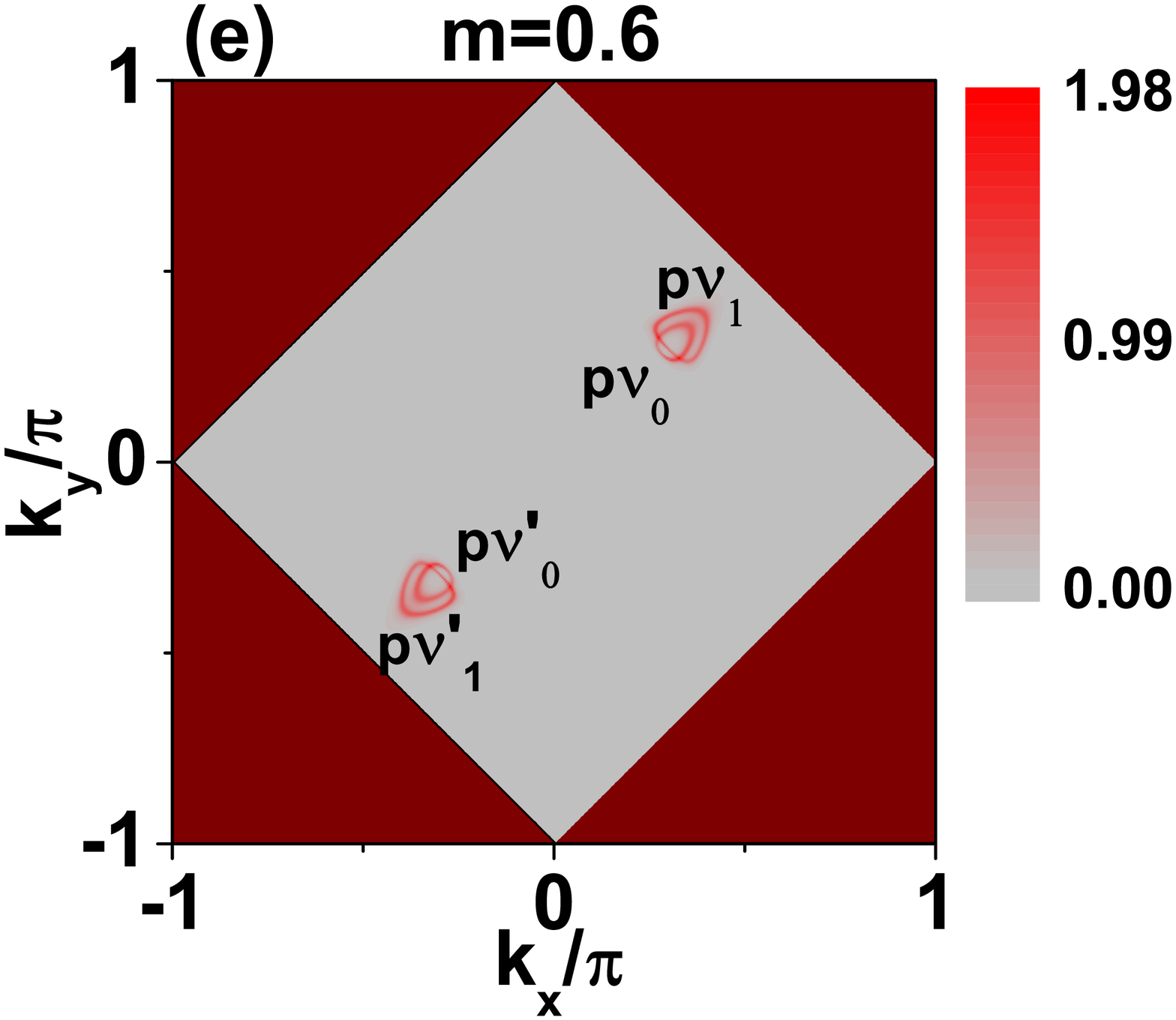}
\caption{(color online) Panel (a) plots the Fermi surface of the
tight-binding model in the first BZ. Panel (b) plots the LDOS of the
tight-binding model. Panel (c) shows the LDOS of $m=0.2,0.6$ without
the impurity. Panel (d) shows the zero temperature spectral function
for $m=0.2$ in the MBZ (denoted by grey color). There are four small
pockets ($P\nu_1,P\nu_0,P\nu^{\prime}_0,P\nu^{\prime}_1$) aligning
along the diagonal direction and two high-intensity squares
($L\nu_0$, $L\nu_1$). Panel (e) is similar to panel (d), but for
$m=0.6$, here only four pockets are left. }\label{fig2}
\end{figure}

The impurity effect in the SDW state depends on the detail of
electronic structure. Since we mainly focus on the low-energy
structures of the LDOS, the FS topology should be important for the
results. The black lines in Figs.~\ref{fig2}(a)
 show the FSs of the tight-binding
model, where the two hole pockets centered around $\Gamma$ point
$(0,0)$ are associated with $\epsilon_{1\nu}$, and the two electron
ones around $M$ point $\pi(\pm1,\pm1)$ are associated with
$\epsilon_{0\nu}$, in which $\nu=0$ (1) represents the inner (outer)
Fermi surfaces of the hole or electron pockets. Without impurity,
LDOS is uniform and site independent, Fig.~\ref{fig2}(b) shows the
LDOS in the normal state, different from it, two peaks show up in
the SDW state. Both the energy gap and the height of the peaks are
increased with the increasing of the value of magnetic order which
can be seen clearly in Fig.~\ref{fig2}(c). In our calculation the
quasipartical damping is $\eta=0.01$. Here we calculate the spectral
function $A(k,\omega)$ at $\omega=0$ in the SDW state without
impurity, which is imaginary part of Green's function multiplied by
$-\frac{1}{\pi}$ and is proportional to the photoemission intensity
measured by ARPES experiment. As can be seen in Fig.~\ref{fig2}(d),
the locations of the bright pockets align along the diagonal
direction which are denoted by
$P\nu_1,P\nu_0,P\nu^{\prime}_0,P\nu^{\prime}_1$, respectively, and
have relations with Dirac cones~\cite{diraccone}. In addition,
although the FSs are mostly be gapped, the gap value is extremely
small around the $\Gamma$ point, thus there are two high-intensity
squares denoted by $L\nu_0$ and $L\nu_1$. While for $m=0.6$, the two
high-intensity squares around the $\Gamma$ point disappear, there
are only the bright spots along the diagonal direction and the
pockets are enlarged compared to the $m=0.2$ case which can be seen
in Fig. \ref{fig2}(e). Actually, our starting model has $C_4$
symmetry when rotating around an As ion, which will be broken by a
single impurity on Fe atom even if there is no SDW order.


\section{Quasiparticle interference for small value of magnetic order}
\begin{figure}
       \includegraphics[width=1.68in]{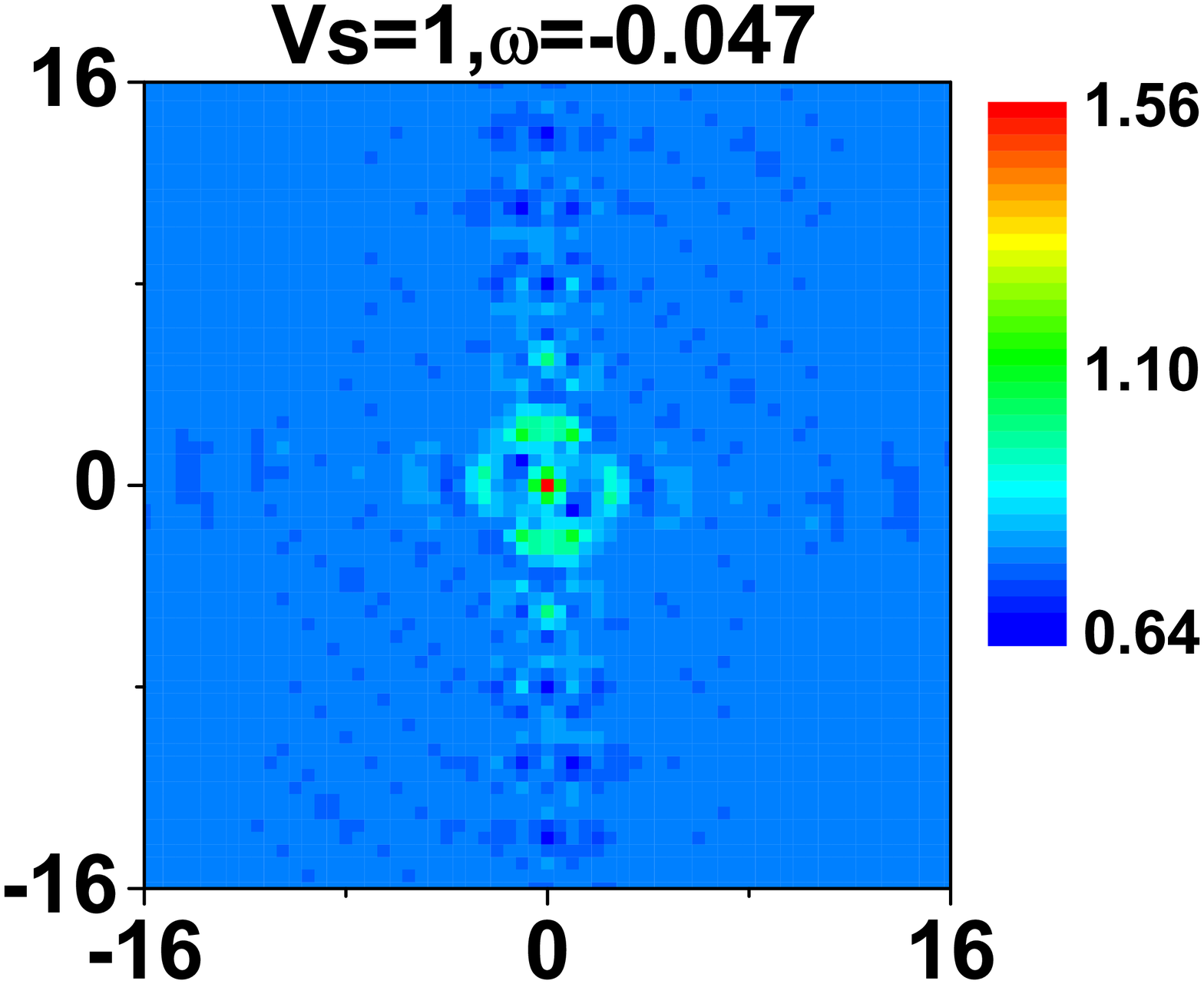}
      \includegraphics[width=1.68in]{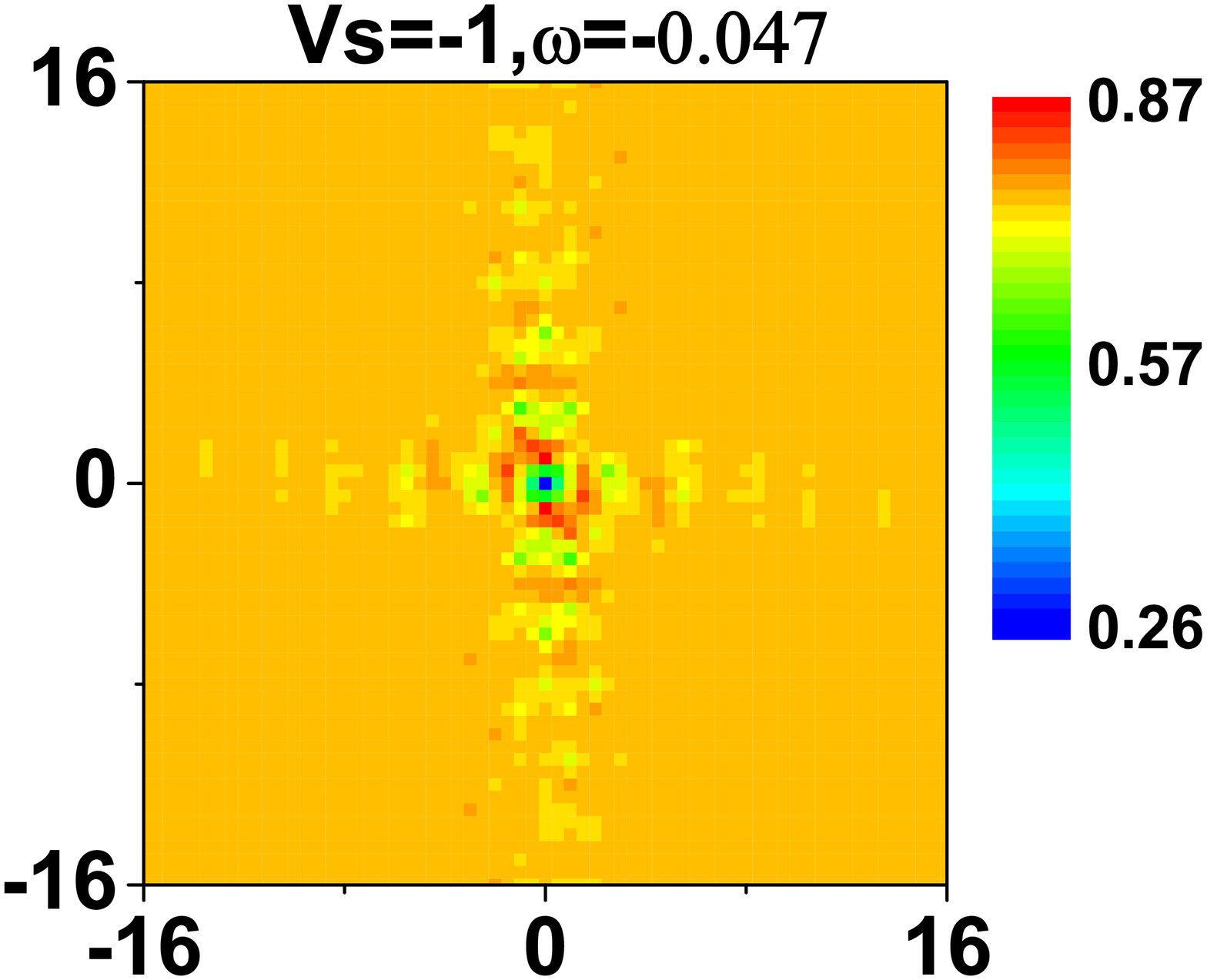}
       \includegraphics[width=1.68in]{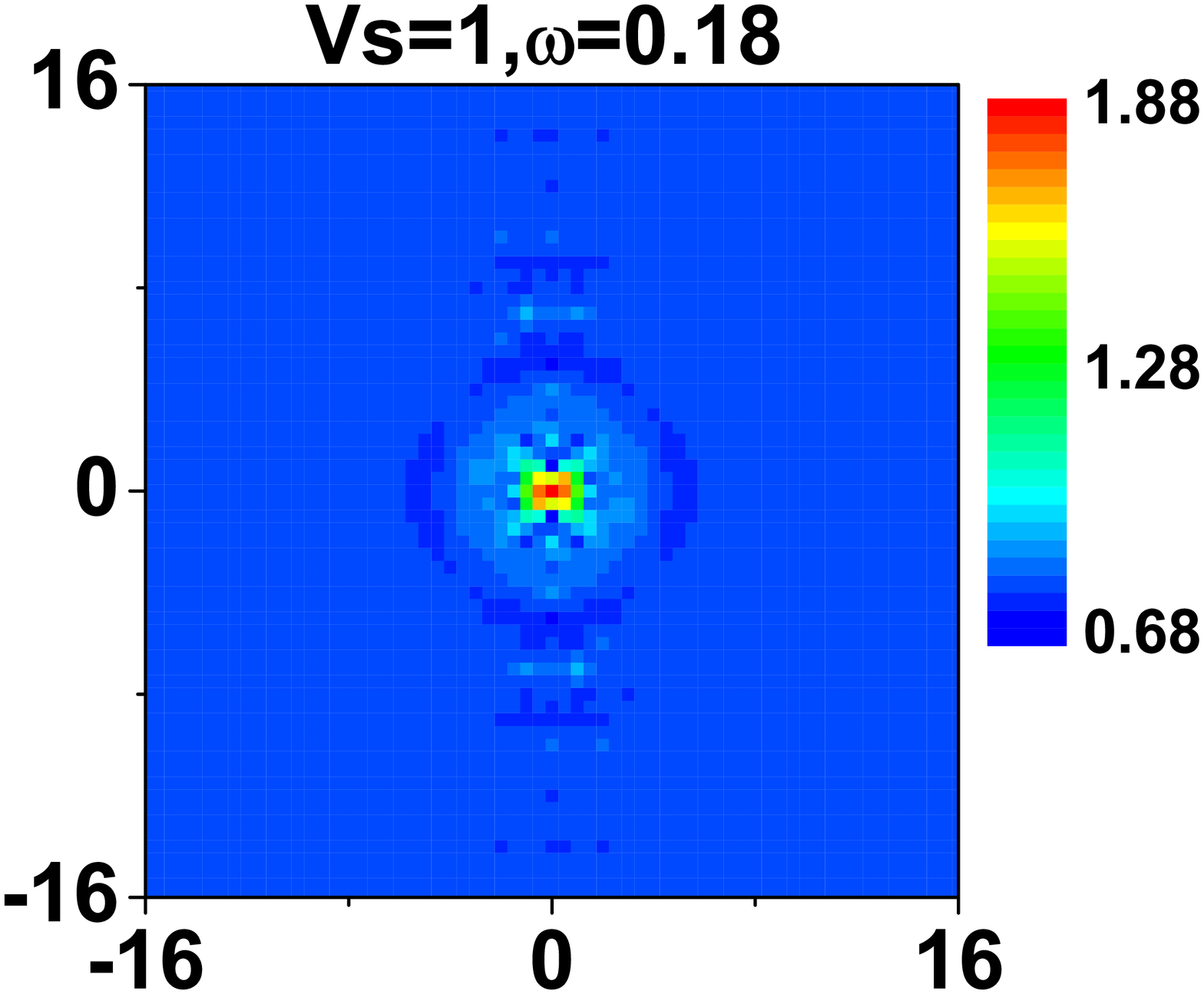}
      \includegraphics[width=1.68in]{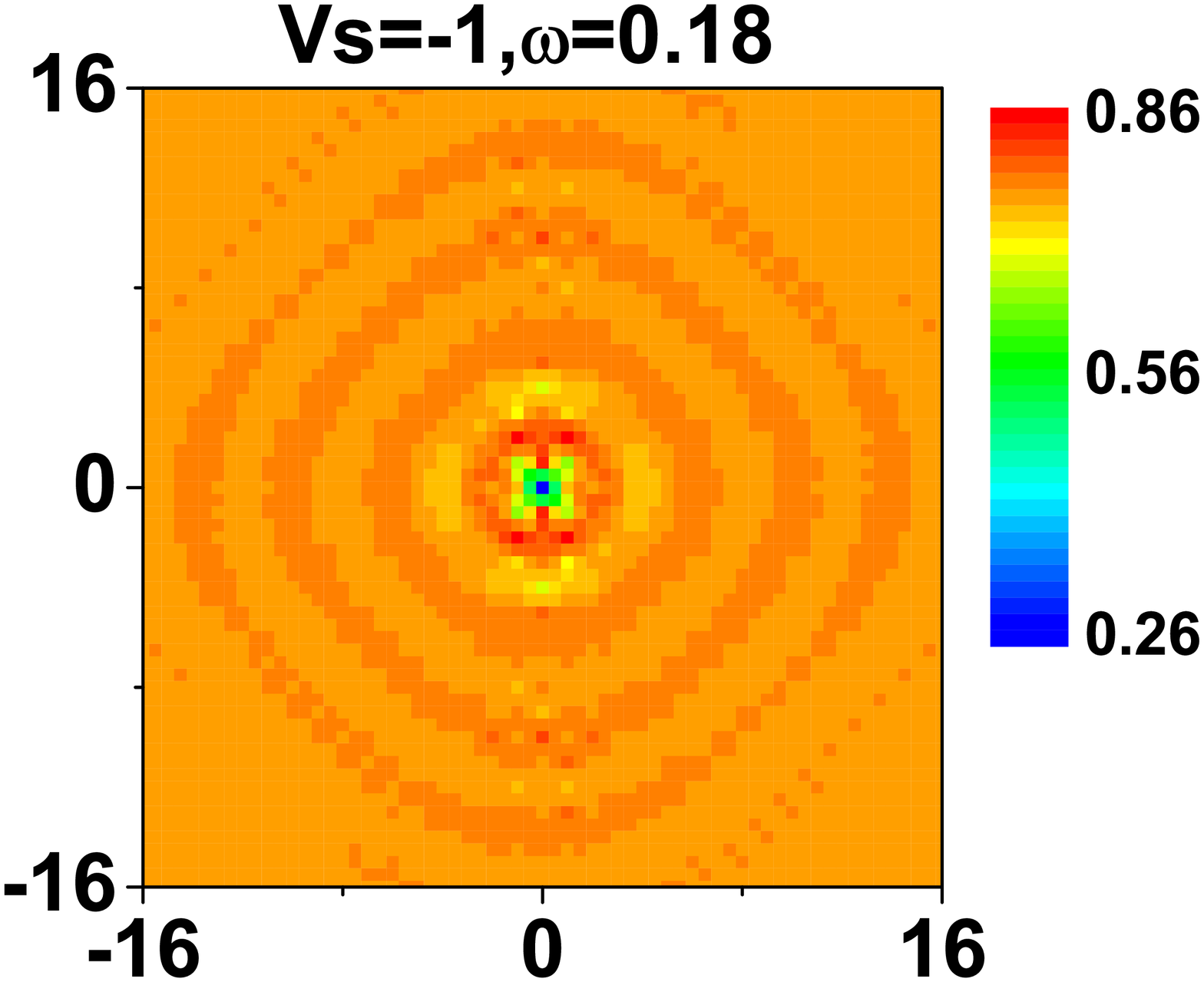}
\caption{(color online)For $m=0.2$, image plot of LDOS in real space
at selected energy for SP $Vs=\pm 1$. The x-axis and y-axis denote
the coordinate of real space.} \label{fig5}
\end{figure}

In the SDW state $m=0.2$, when the SP is weak, the LDOS on the
impurity site has finite value. Due to the scattering of impurity,
LDOS is site dependent. Spatial modulations of LDOS at energies
$-0.047$ and $0.18$ are shown in Fig.~\ref{fig5} for $Vs=\pm
1$.there are $33\times33$ sites in sublattice A and $32\times32$
sites in sublattice B. Those energies corresponding to the two SDW
peaks of LDOS for $Vs=1$ at the impurity site which we do not show
here. Intensity of LDOS is enhanced at the impurity site for $Vs=1$.
On the contrary, it is suppressed for $Vs=-1$. We can see that
modulations exist along x-axis as well as along y-axis. In the case
of $Vs=-1,\omega=0.18$, the QPI exhibits a 2D ripple-like
modulation, the wave length is about $4a$.

\begin{figure}
      \includegraphics[width=1.49in]{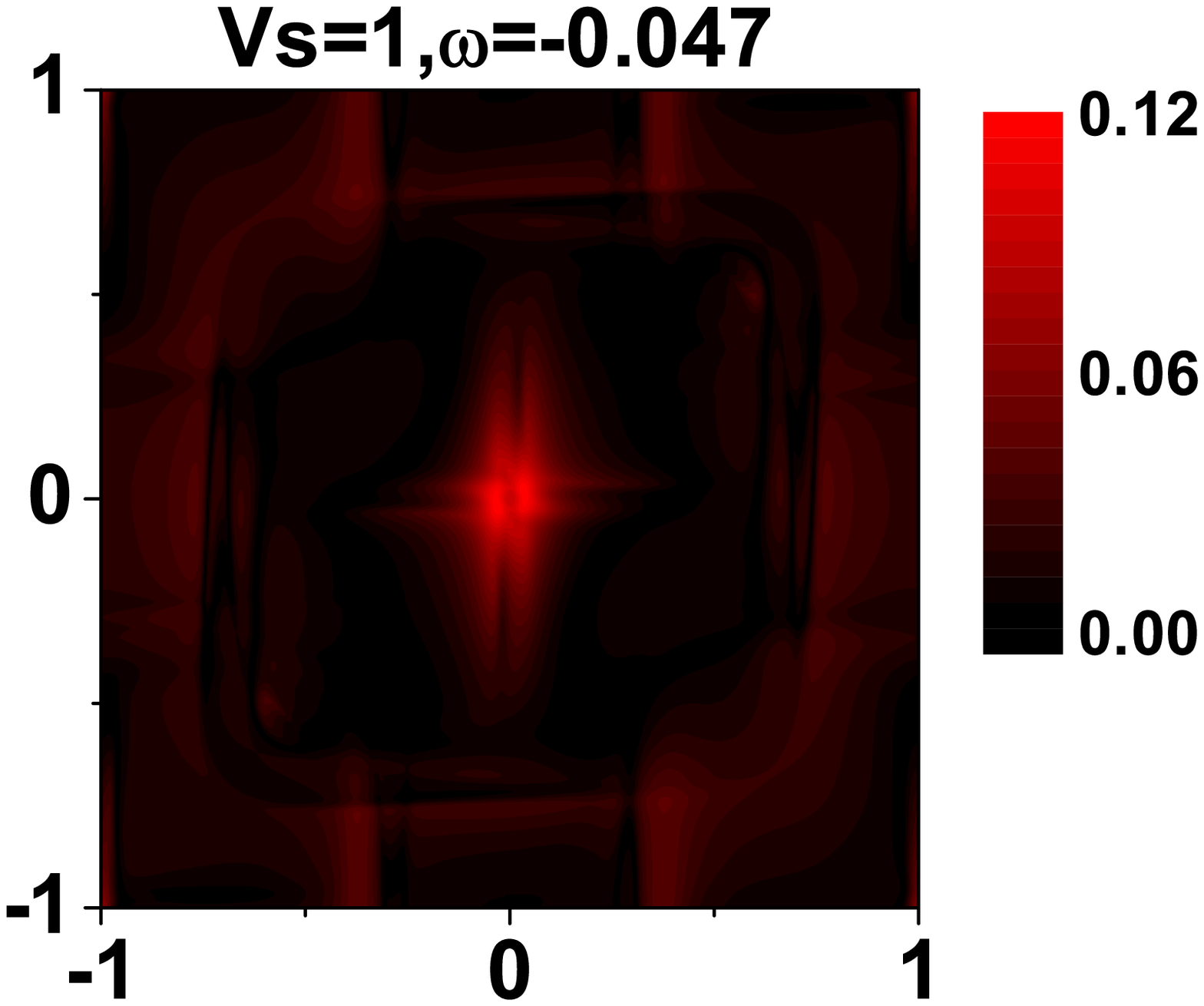}
      \includegraphics[width=1.68in]{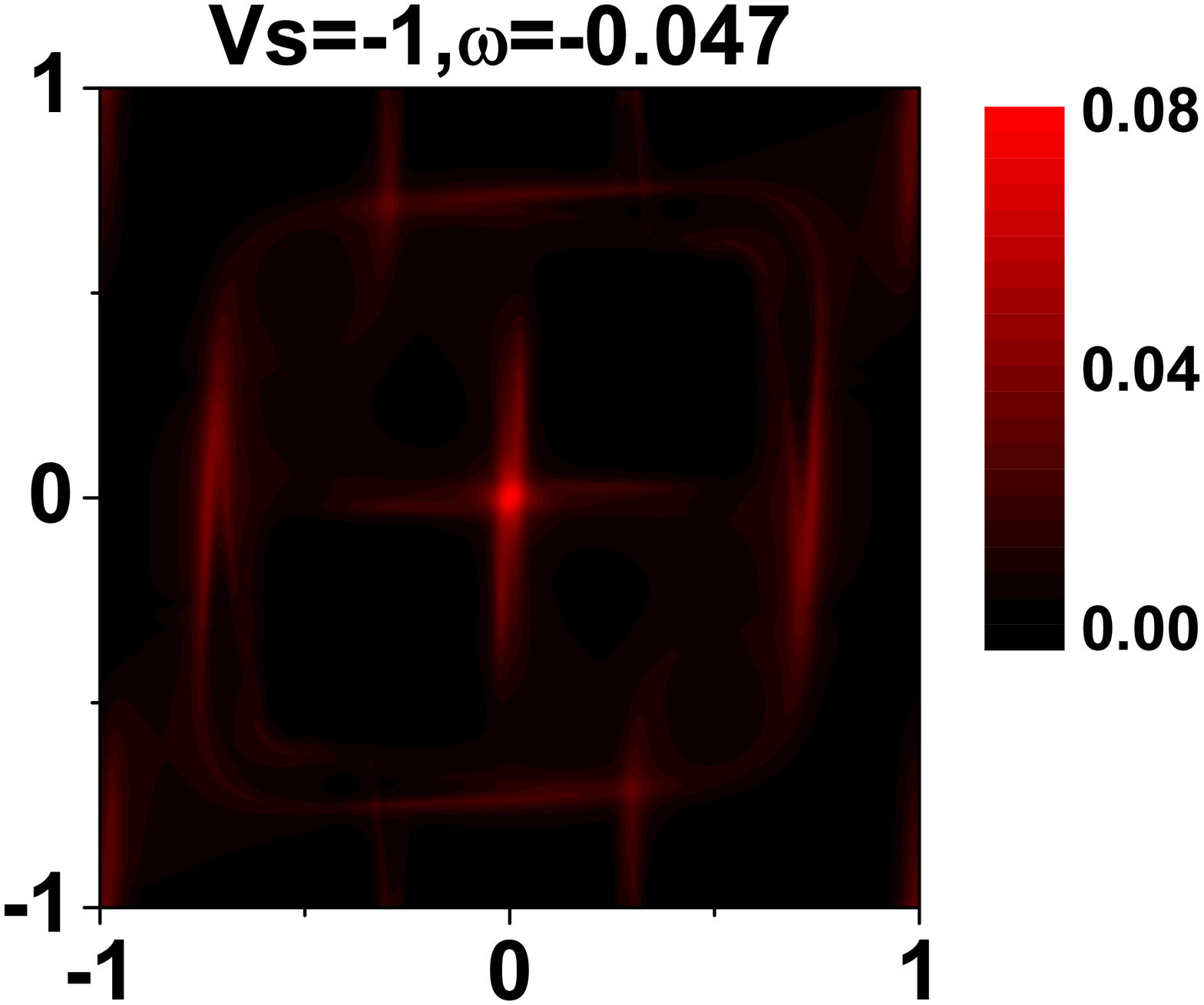}
      \includegraphics[width=1.68in]{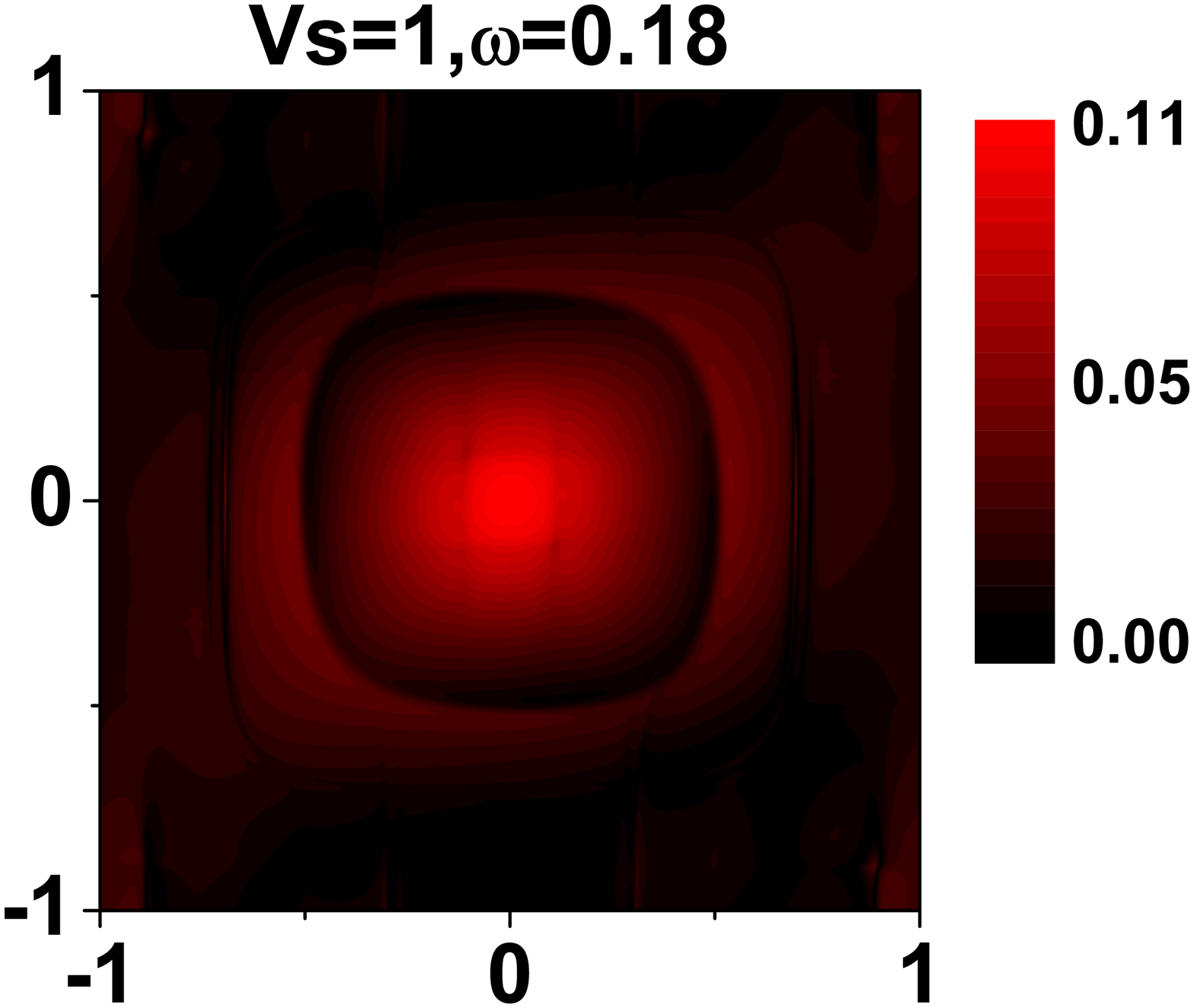}
      \includegraphics[width=1.68in]{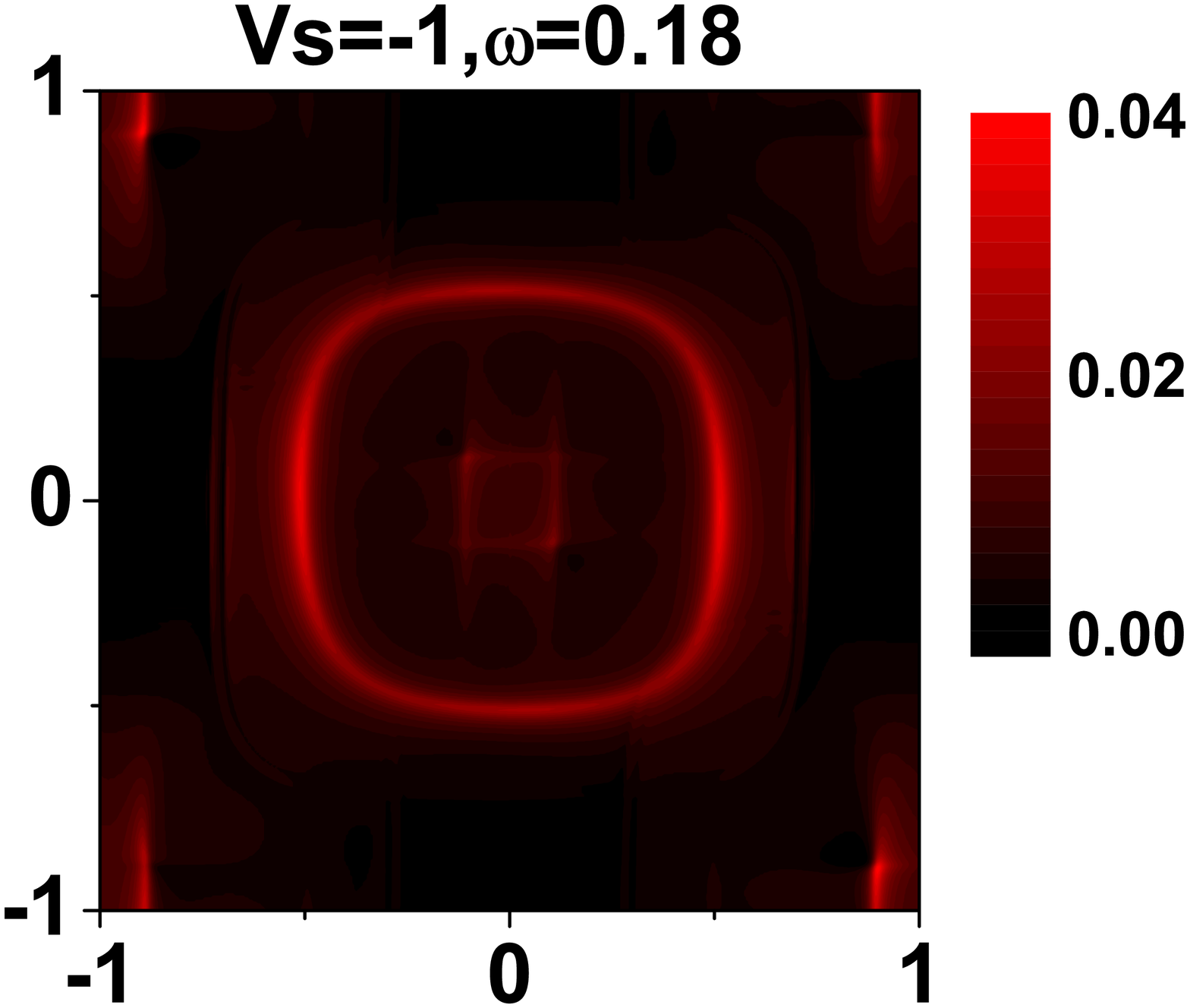}
\caption{(color online)The FC-LDOS $\rho_q(\omega)$  for the $m=0.2$
 are shown in panels at selected energy for $Vs=\pm1$ in the
first BZ. The unit of the bar is $10^{-4}$.} \label{fig6}
\end{figure}

Then we plot the image map of the FC-LDOS in the SDW state at
selected energies in Fig.~\ref{fig6}. We can see that in q space,
there appear high intensity lines, the corresponding wave vectors
are responsible for the QPI. The interference is nearly equally
strong along x-axis and y-axis. At energy $\omega=-0.047$, for SP
$Vs=\pm1$, the high intensity lines near the center are due to the
scattering of pockets($P\nu_{i}$,$P\nu^{\prime}_{i}$) to the
corresponding squares($L\nu_{i}$). As seen from Fig.~\ref{fig6}, for
$Vs=-1,\omega=0.18$, the value of the scattering $q$ along the
circle is about $|q|=0.5\pi$, consistent with the wave length $4a$
in real space. It also indicates that when bias energy deviate from
zero, the underlying band structure is very important.


\begin{figure}
      \includegraphics[width=1.68in]{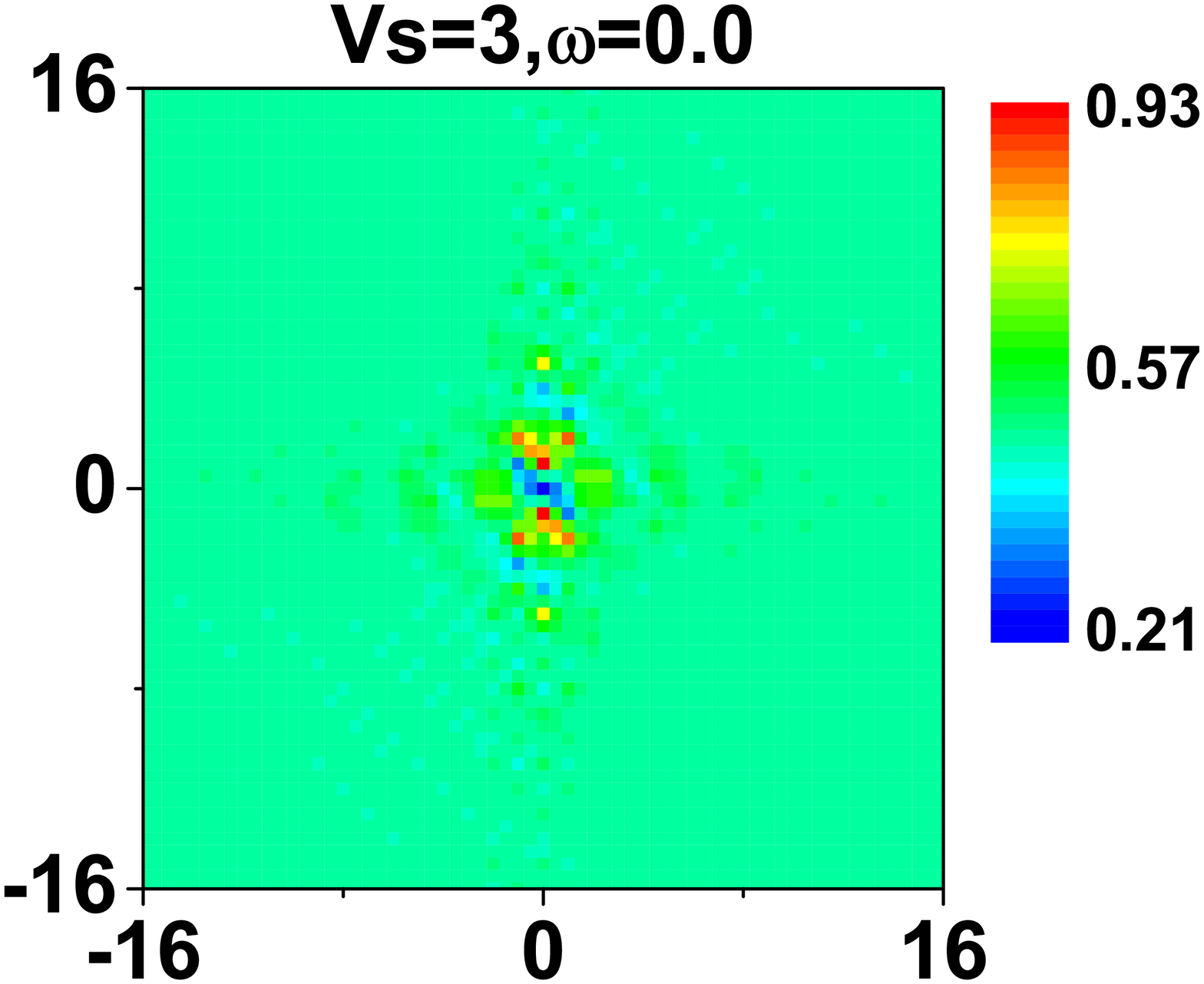}
      \includegraphics[width=1.68in]{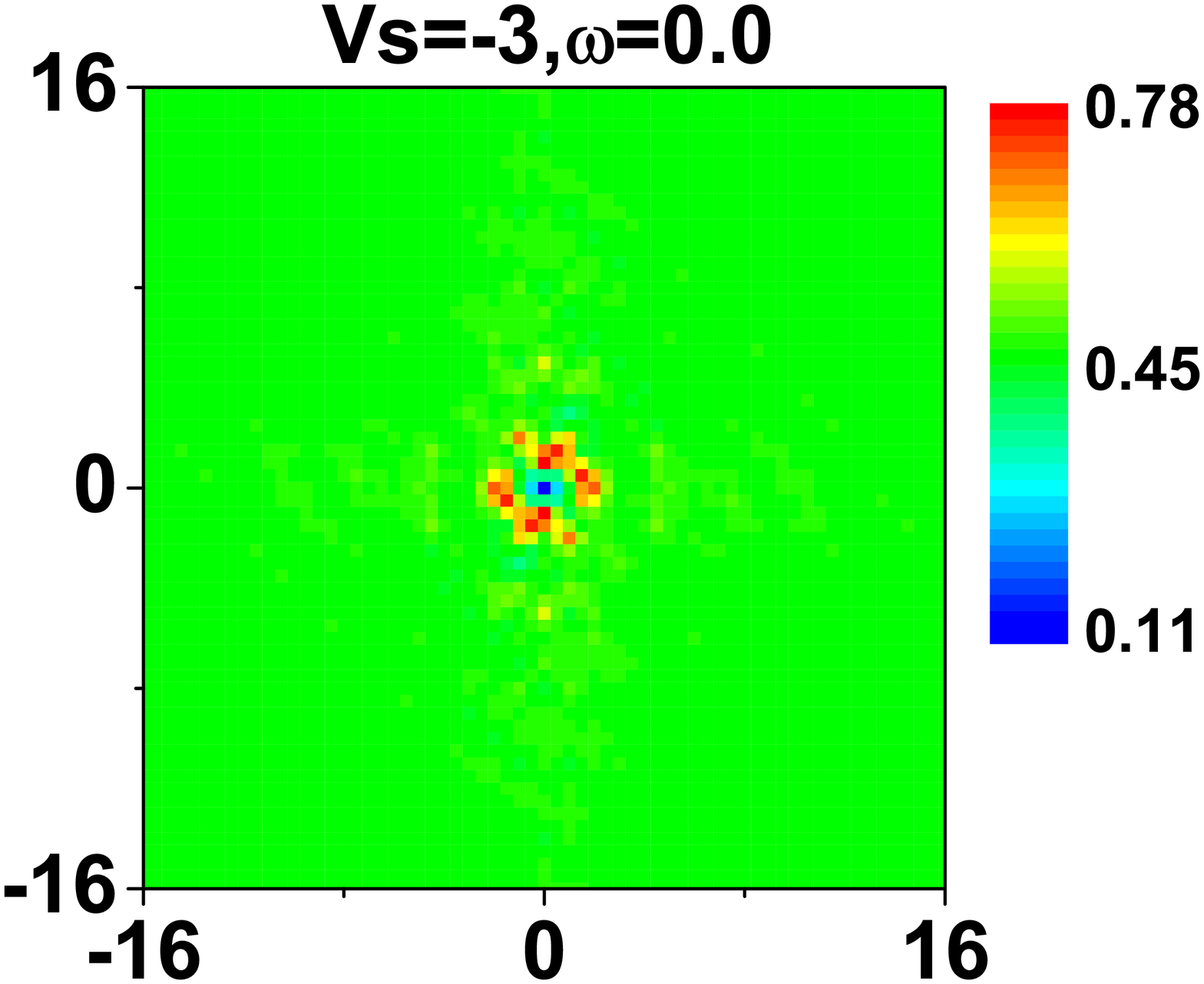}
      \includegraphics[width=1.68in]{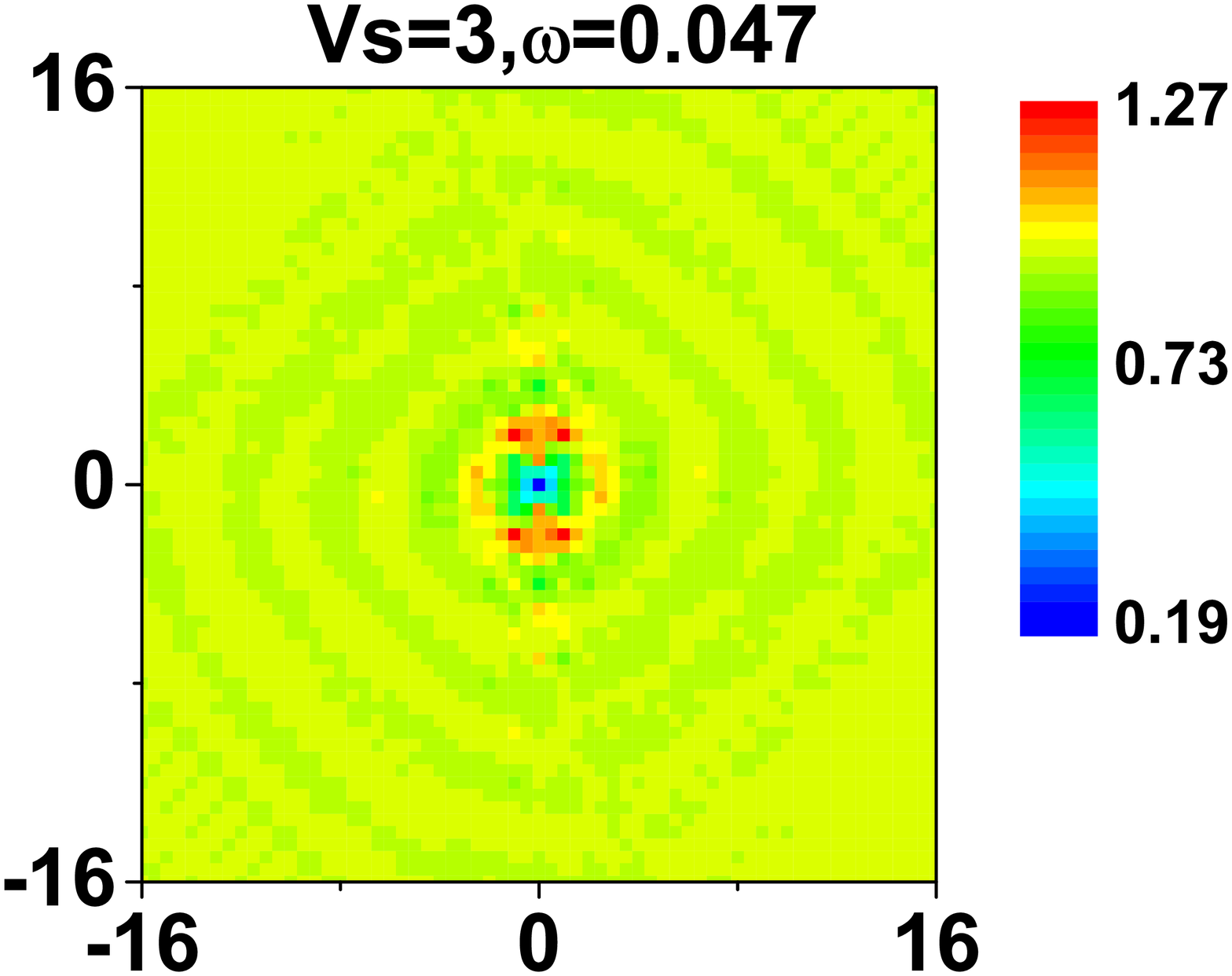}
      \includegraphics[width=1.68in]{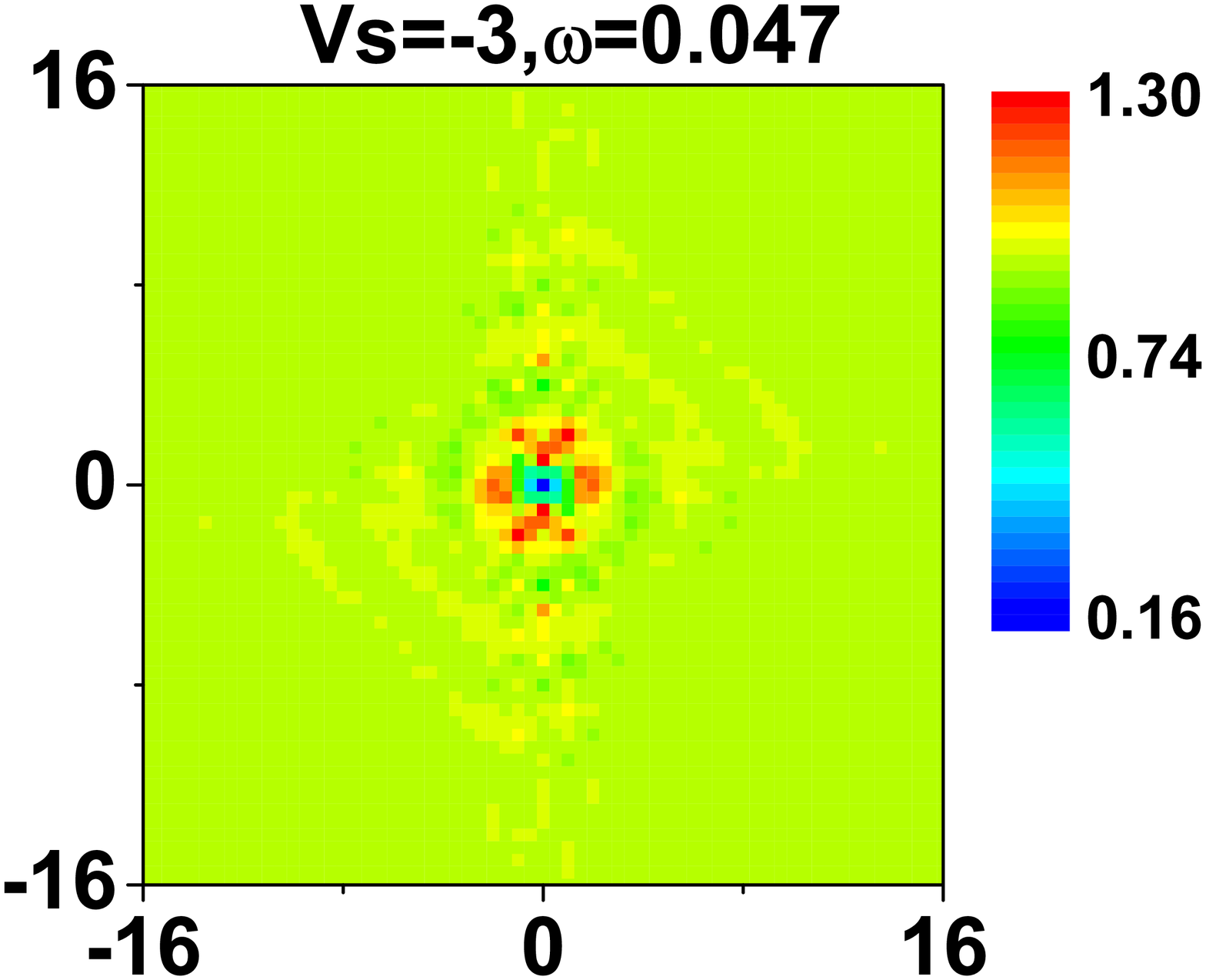}
\caption{(color online)Similar to Fig.\ref{fig5}, but for SP $Vs=\pm
3$.} \label{fig8}
\end{figure}
\begin{figure}
      \includegraphics[width=1.56in]{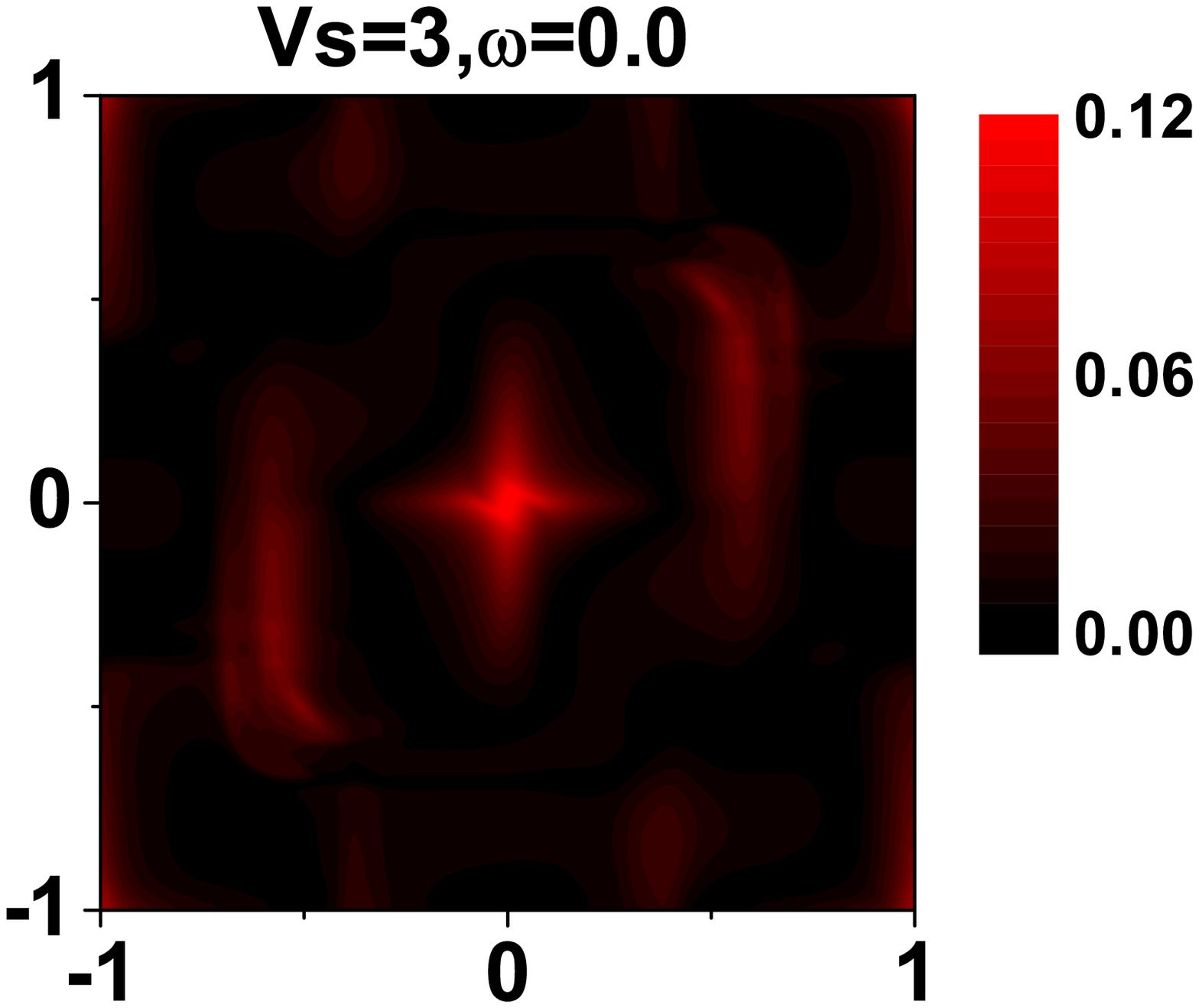}
      \includegraphics[width=1.56in]{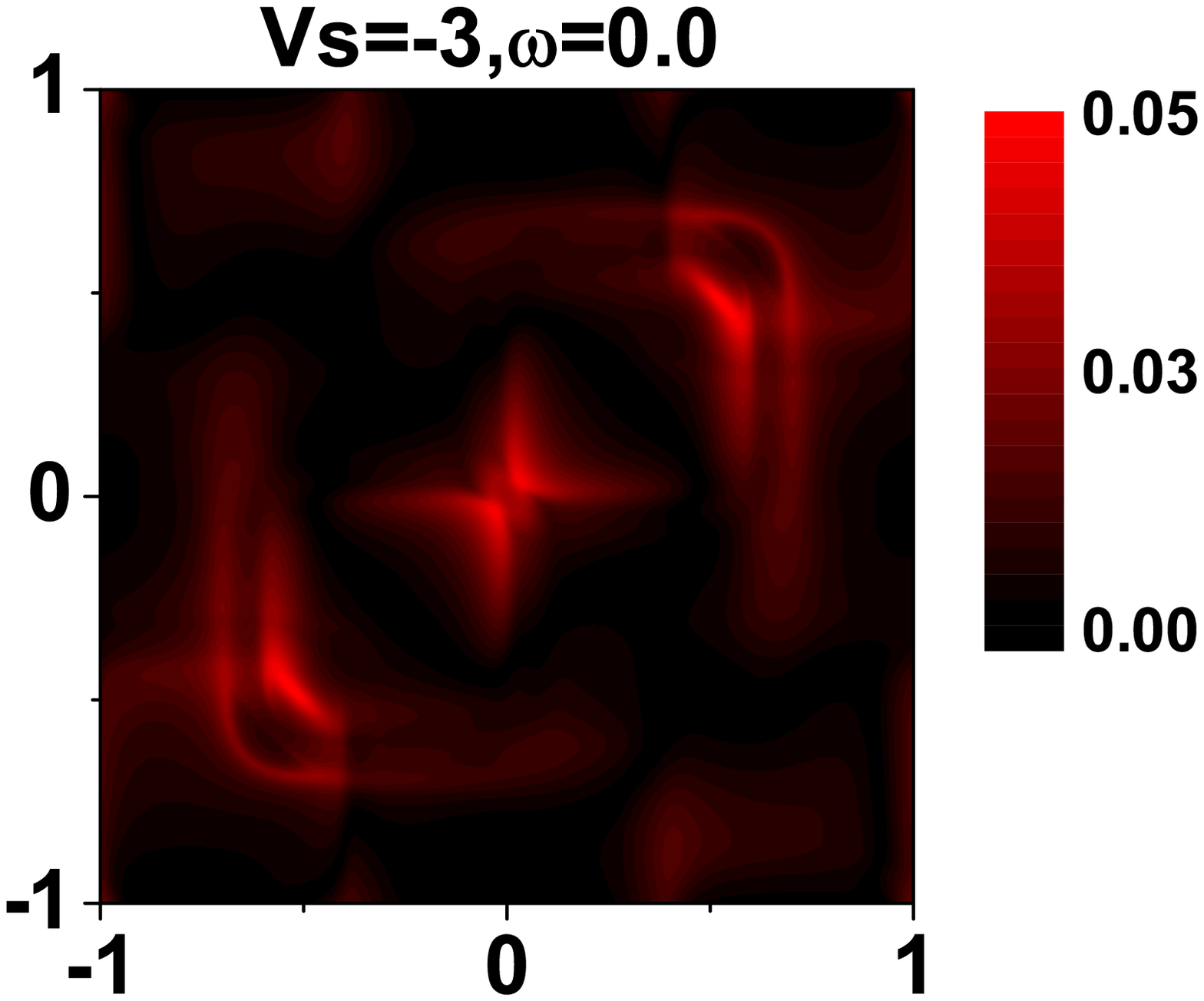}
      \includegraphics[width=1.56in]{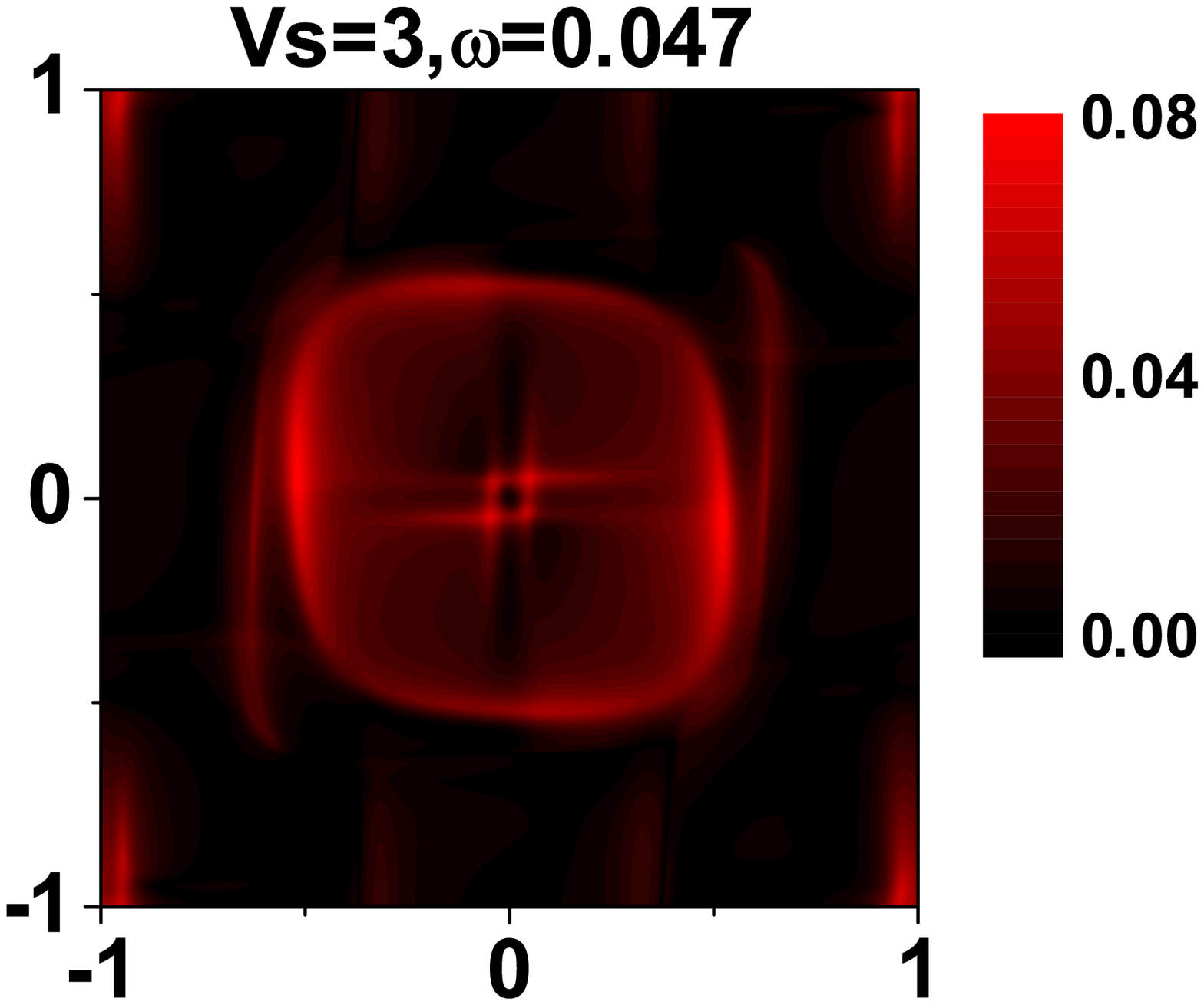}
      \includegraphics[width=1.56in]{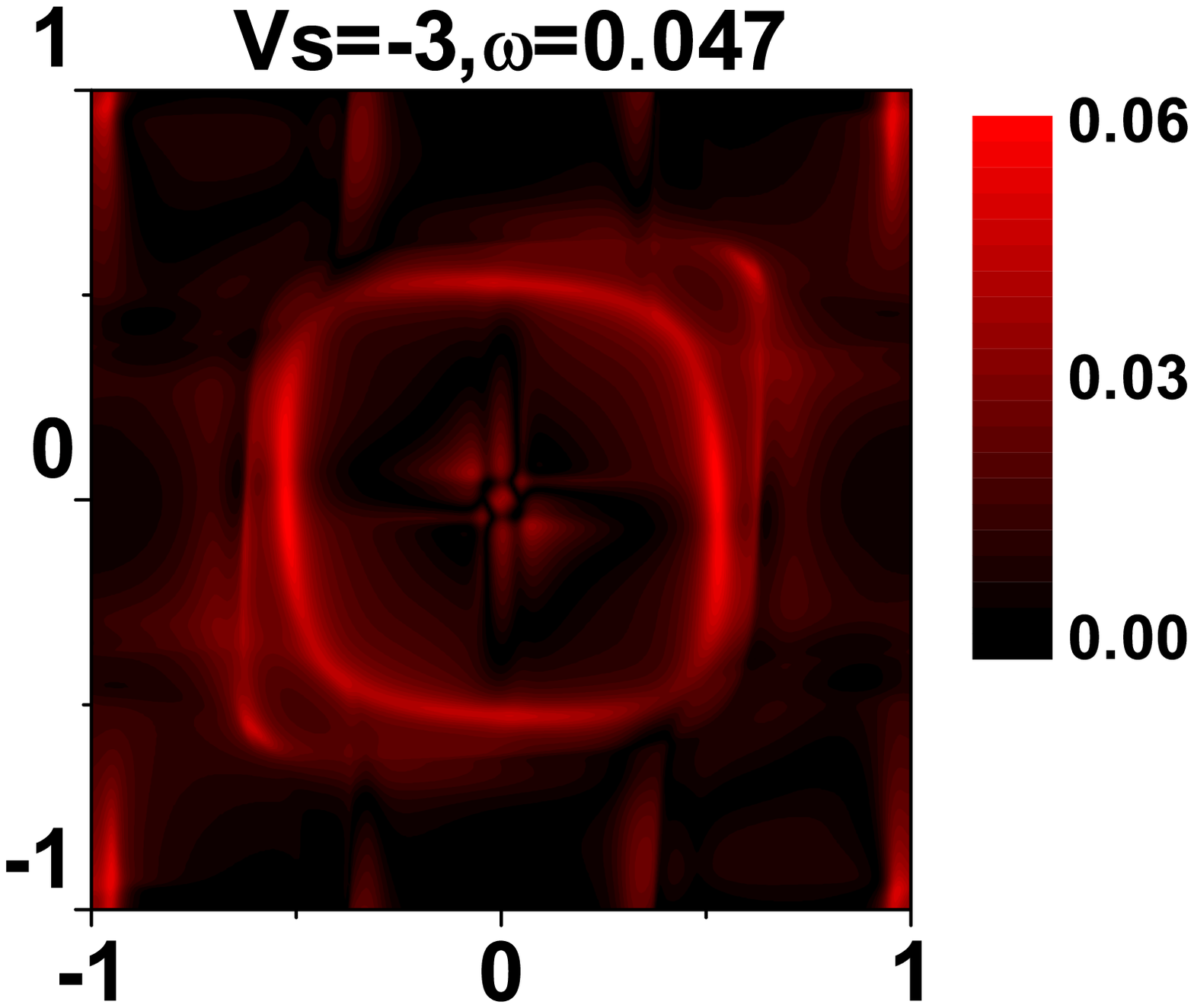}
\caption{(color online)Similar to Fig.\ref{fig6}, but for SP $Vs=\pm
3$.} \label{fig9}
\end{figure}

As SP increases to $Vs=\pm 3$, the LDOS at the impurity site
decreases rapidly and will vanish for larger SP.
Fig.~\ref{fig8} shows that the modulation of the LDOS still exhibits
$C_2$ symmetry. At $\omega=0.047$, 2D ripple-like pattern appears.
From the $q$ space map of $\rho_q(\omega)$ in Fig.~\ref{fig9}, we
note that for strong SP $Vs=\pm3 $, at the Fermi energy, the
intra-pocket scattering leads to the two high intensity small arcs
near the center. While the scattering between $L\nu_{0}(L\nu_{1})$
and $P\nu_{0},P\nu^{\prime}_{0}(P\nu_{1},P\nu^{\prime}_{1})$  lead
to the off-diagonal high intensity spots. For $Vs=-3,\omega=0.0$,
away from the center, along the diagonal direction the high
intensity arcs arise from inter-pocket scattering(from $P\nu_{i}$ to
$P\nu^{\prime}_{i}$) which can be seen clearly from Fig.~\ref{fig9}.
At $\omega=0.047$, the hight intensity wave vectors along the circle
are responsible for the ripple-like modulation in real space. The
interplay of FS
 with the underlying band structure has crucial effect in the
scattering process, since the high intensity spots are obviously
related to the band structure. For more strong SP, the feature of
LDOS and the modulation of LDOS is similar, we also note that for
large value of SP, the difference between the repulsive and
attractive potentials becomes less obvious.

\section{Quasiparticle interference for larger value of magnetic order}
\begin{figure}
      \includegraphics[width=1.68in]{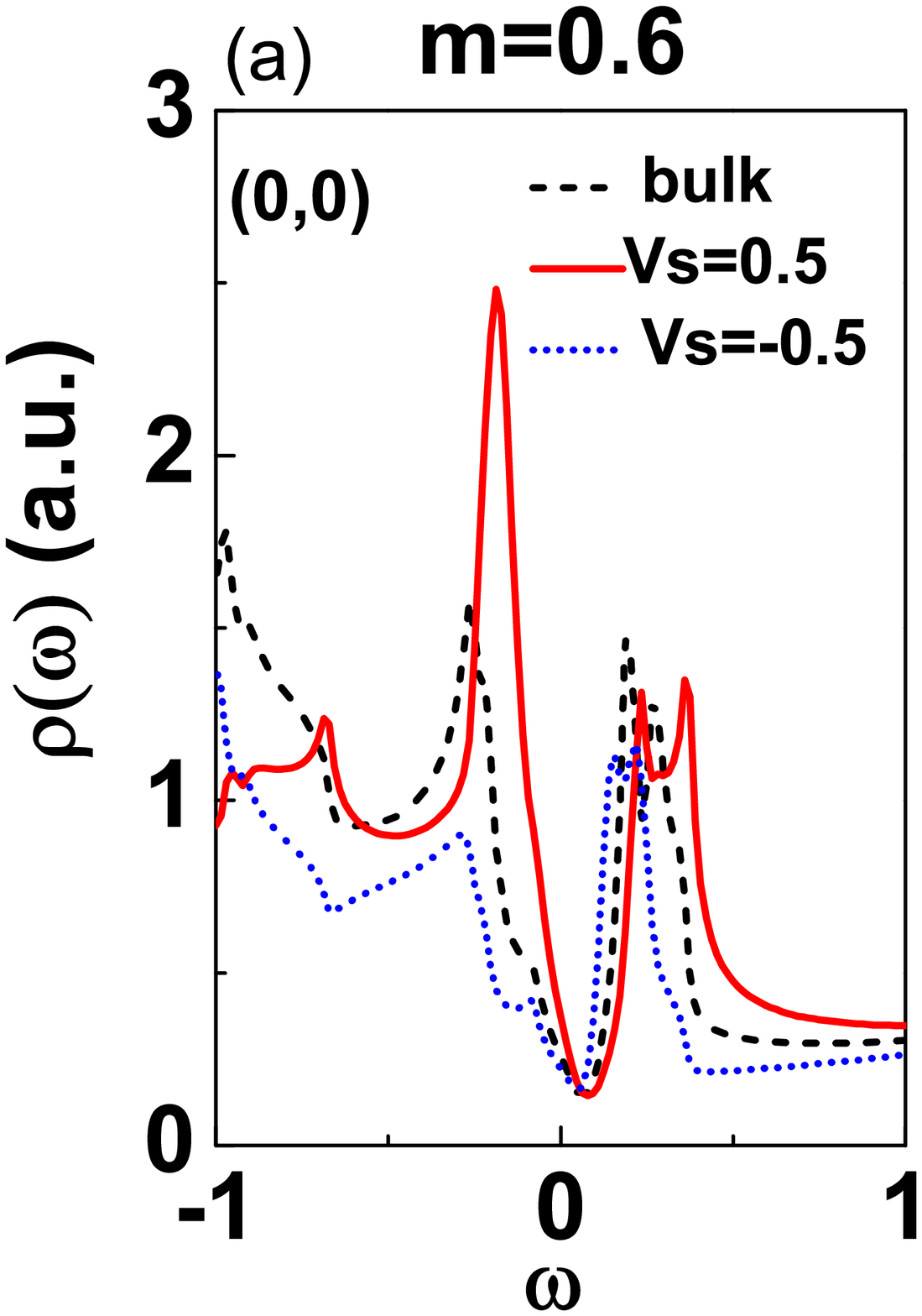}
      \includegraphics[width=1.68in]{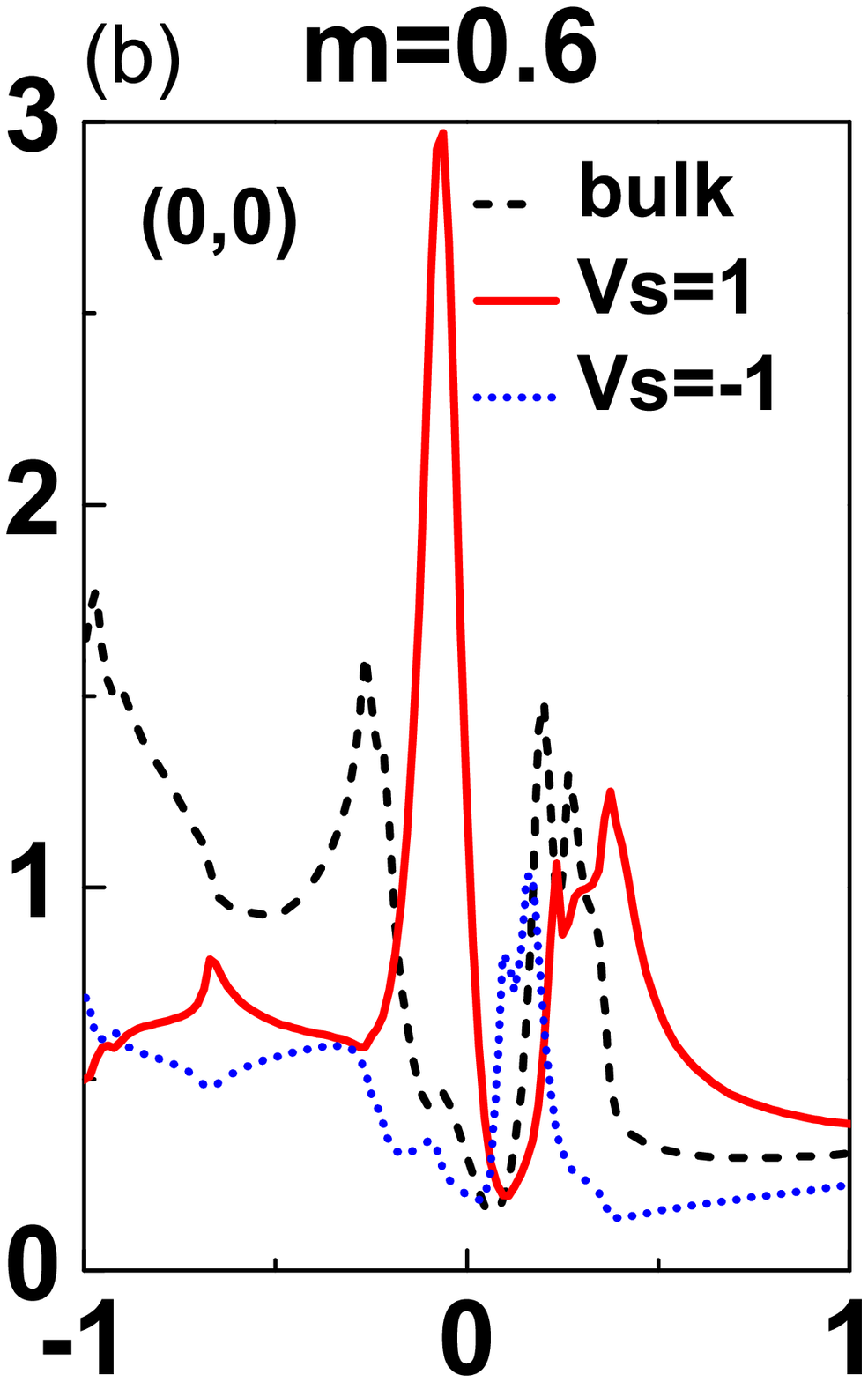}
\caption{(color online)Panel(a) shows LDOS at the impurity site
$(0,0)$ for different scattering potential $Vs=\pm 0.5$ with
$m=0.6$. Black dashed line represents LDOS of the bulk, the red
solid(blue dot) line corresponds to positive(negative) SP. Panel(b)
is similar to (a) with $Vs=\pm 1$. } \label{fig10}
\end{figure}

\begin{figure}
      \includegraphics[width=1.68in]{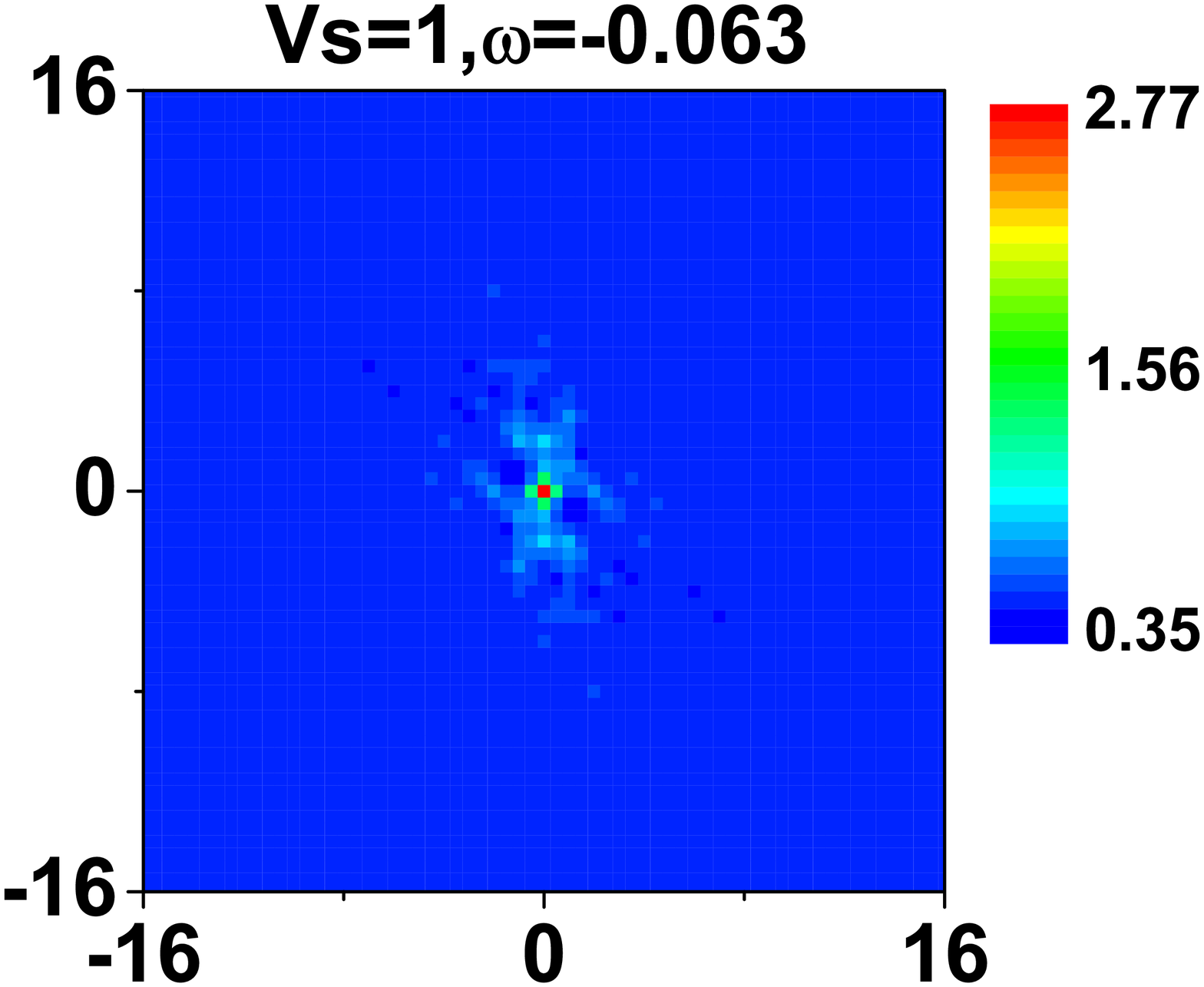}
      \includegraphics[width=1.68in]{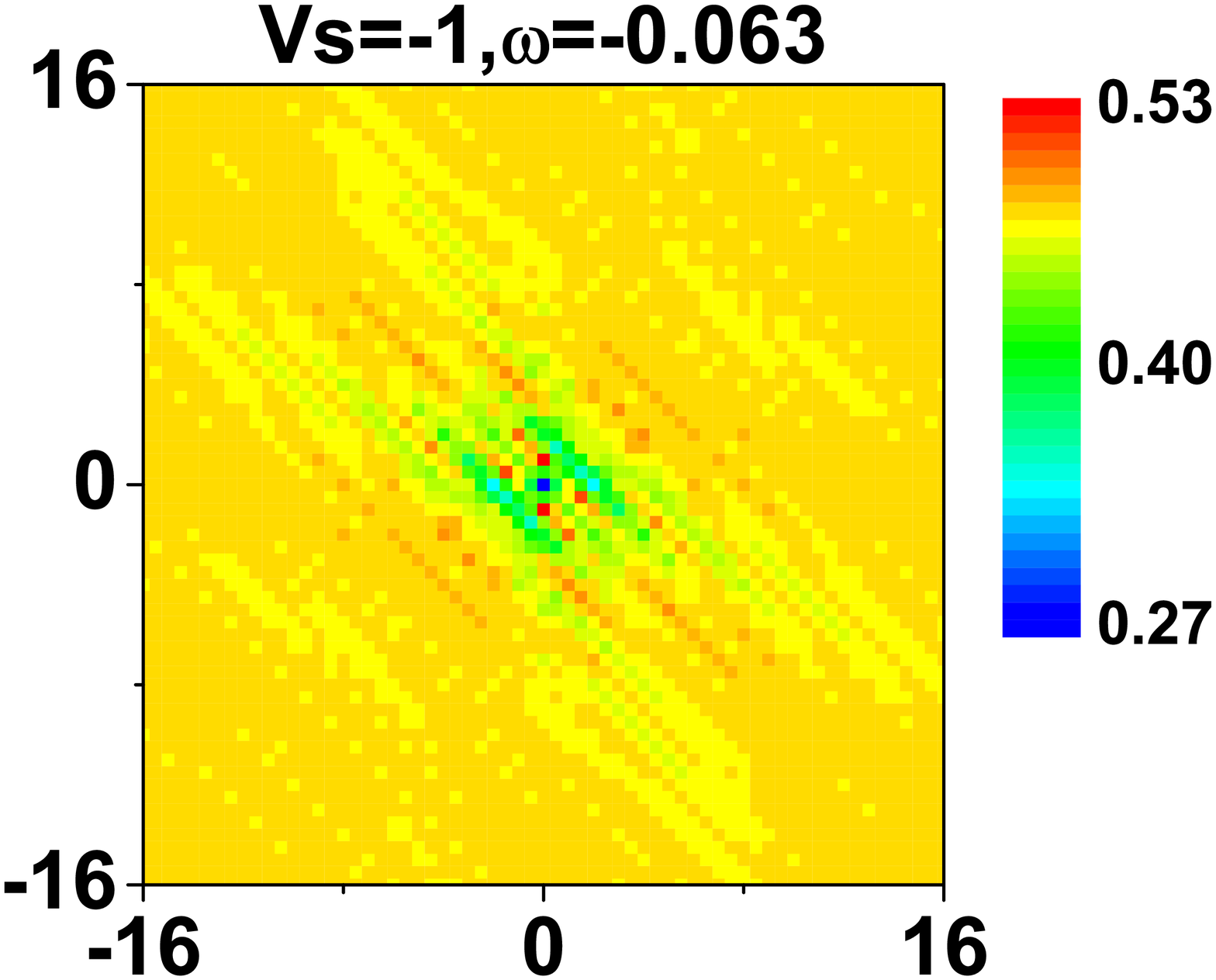}
\caption{(color online)Similar to Fig.\ref{fig5}, but with $m=0.6$.
} \label{fig11}
\end{figure}

Previous discussions show that the value of magnetic order has high
influence on the spectral function, thus we expect it will affect
the QPI as well. For larger magnetic order $m=0.6$ and weak SP, at
the impurity site, the asymmetry of the LDOS is remarkable. For
positive SP $Vs=0.5$, the negative energy peak of the LDOS is much
higher than the positive one. While for $Vs=1$ the resonance peak is
enhanced and pushed to $\omega=-0.063$, near the Fermi energy.
Fig.~\ref{fig10} shows it clearly. On the contrary, for negative SP,
the intensity of the overall LDOS is relatively small and the
positive energy peak is higher. The most striking feature of the
larger-$m$ system is the existence of 1D modulation of the LDOS.
Compared to the positive $Vs=1$ case, the 1D structure is more
remarkable for $Vs=-1$, as can be seen in Fig.~\ref{fig11}. For
$Vs=-1$, at selected energies
$-0.063$, the 1D
stripe pattern is pronounced. The existence of 1D stripe is
consistent with the nematic electronic structure observed in the
parent compound of the 122 systems~\cite{chuang} which have large
magnetic moment.

\begin{figure}
   \includegraphics[width=1.48in]{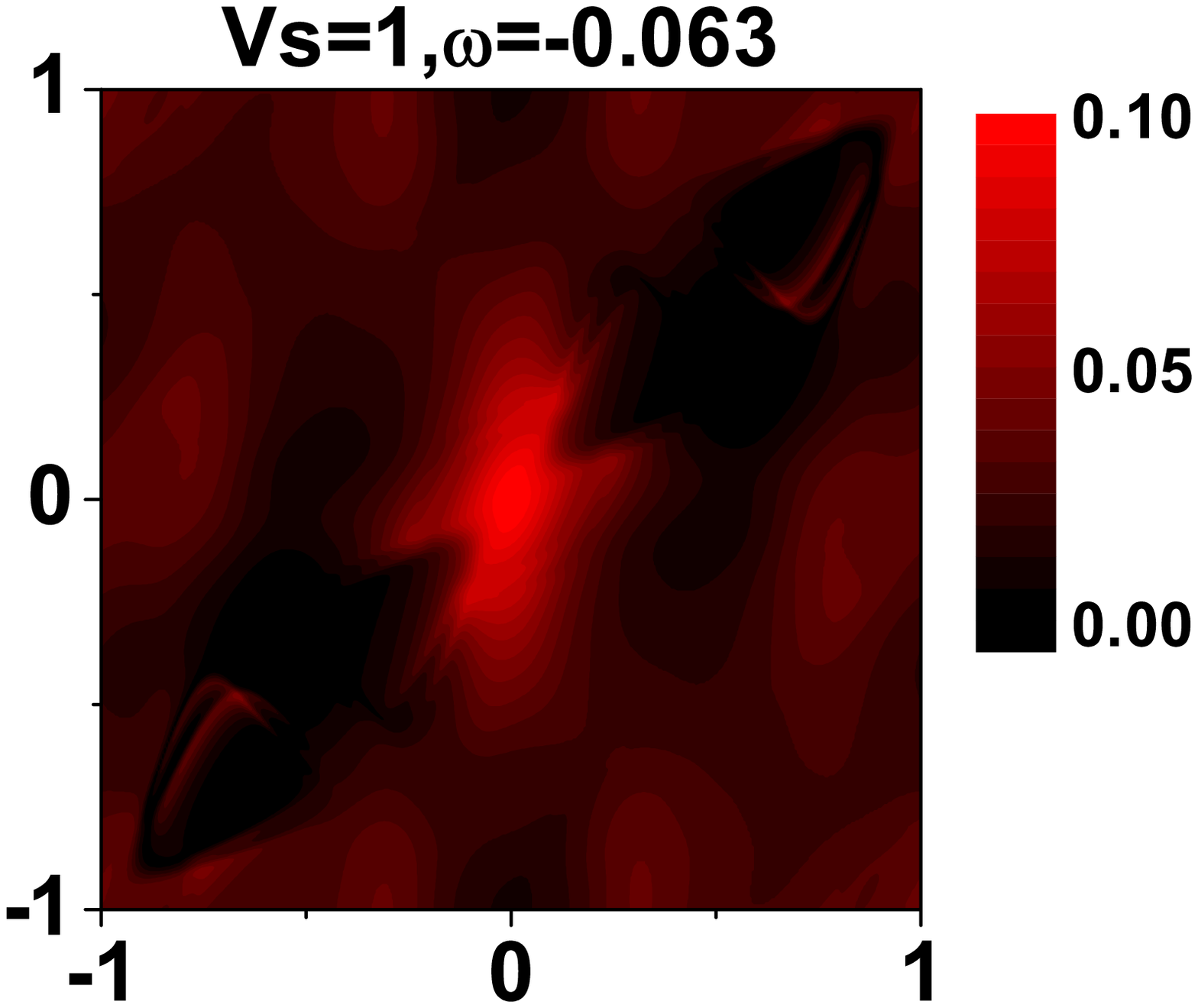}
      \includegraphics[width=1.68in]{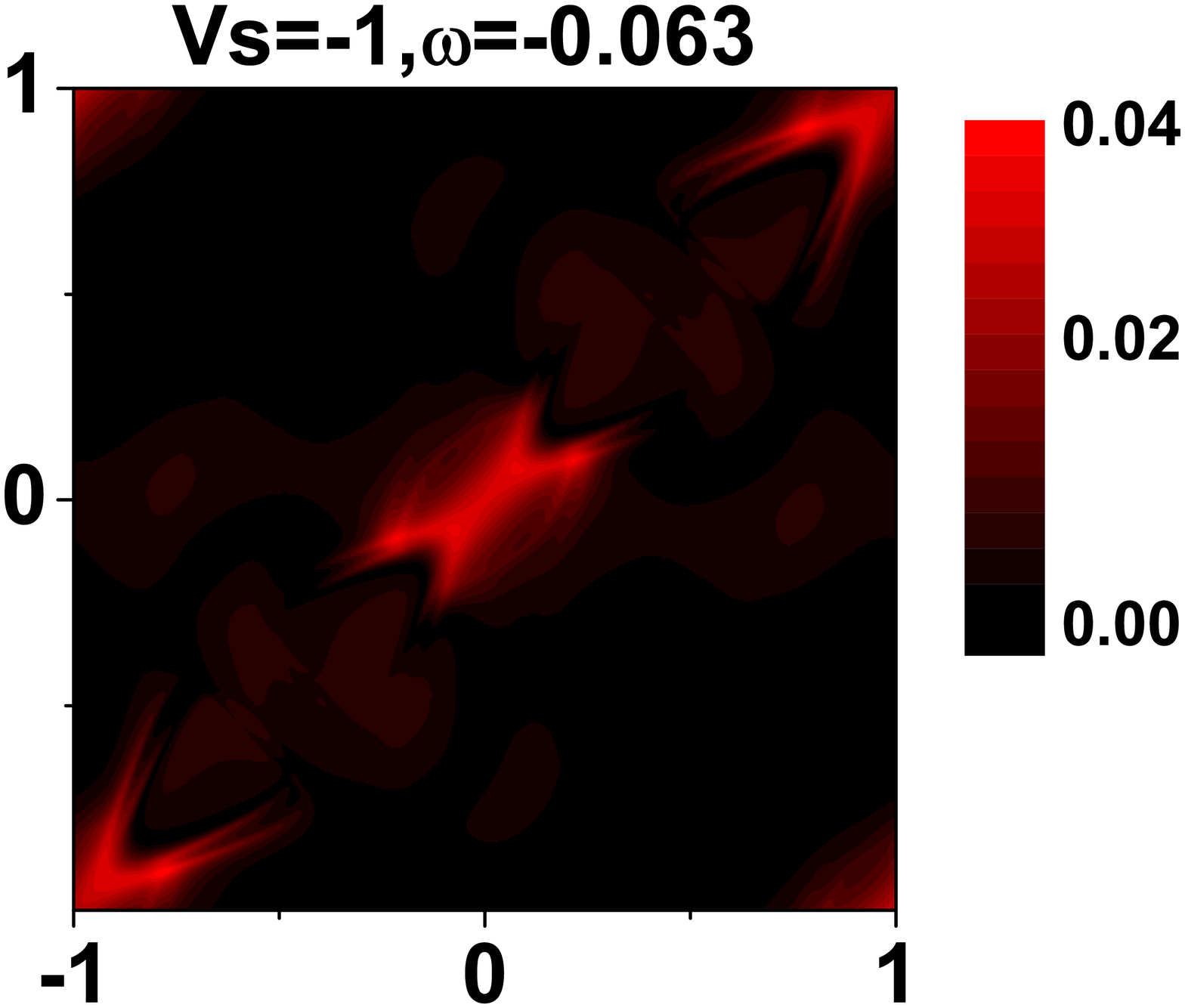}
\caption{(color online)Similar to Fig.\ref{fig6}, but with $m=0.6$.
} \label{fig12}
\end{figure}

For $m=0.6$, only four pockets are left in the plot of spectral
function, which play a important role in the formation of stripe
patterns. Responding to the modulation of the LDOS in real space,
dispersive excitation in q-space should appear along its
perpendicular direction. This is illustrated in Fig.~\ref{fig12}, we
can see that the dominant high-intensity spots are distributed along
a diagonal direction with differently detailed patterns at different
energies. For $Vs=-1$,
at $\omega=-0.063$, the scattering vector $q\simeq(\pm\pi,\pm\pi)$.
They are corresponding to the distance of $\sim2a$ strip pattern in
real space. While the intra-pocket scattering leads to the high
intensity spots around the center. Except for the high intensity
spots around the center, the high-intensity spots for $Vs=-1$
correspond to the dark ones for $Vs=1$, it means that for repulsive
scattering inter-pockets scattering is weak.

\begin{figure}
      \includegraphics[width=1.68in]{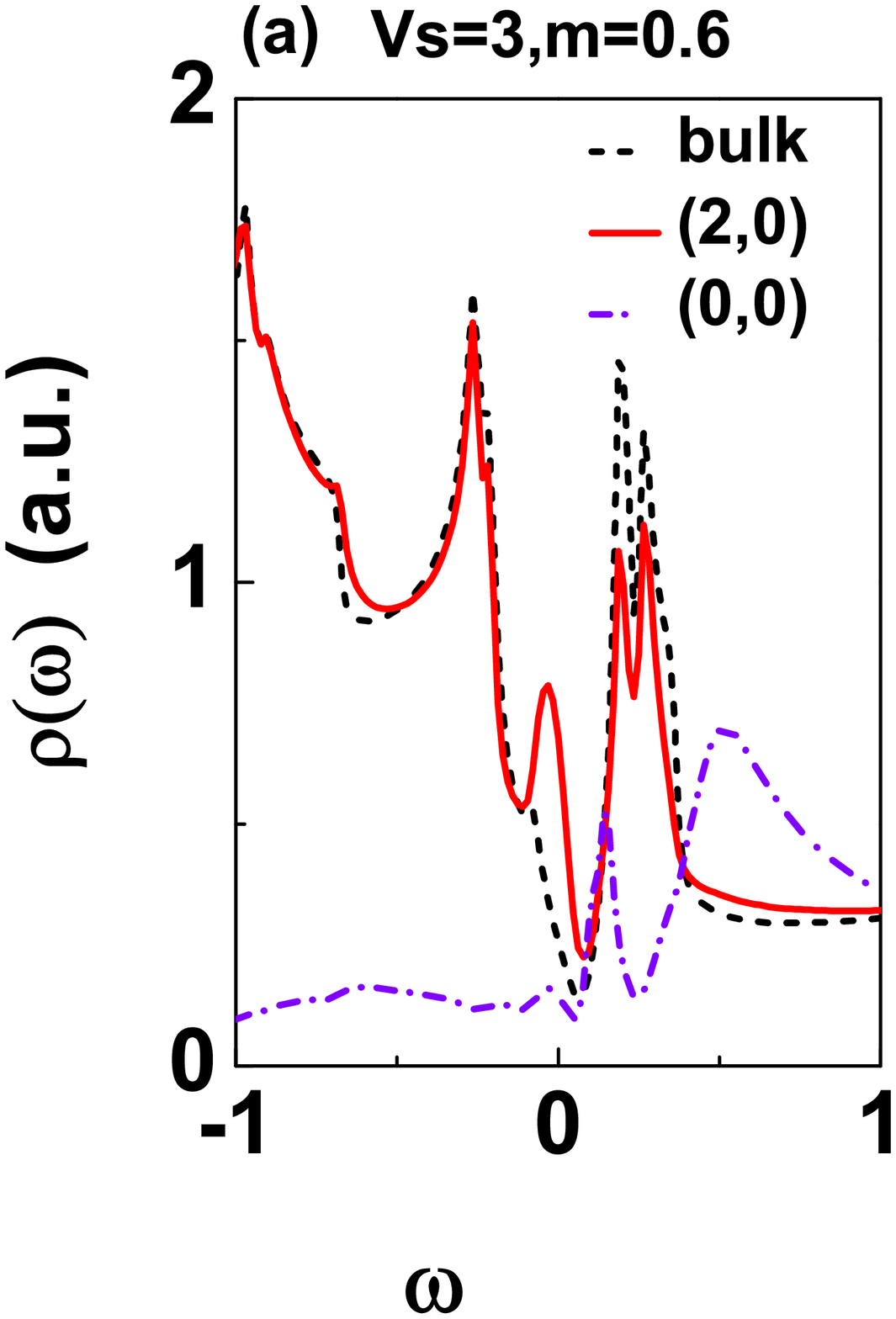}
      \includegraphics[width=1.68in]{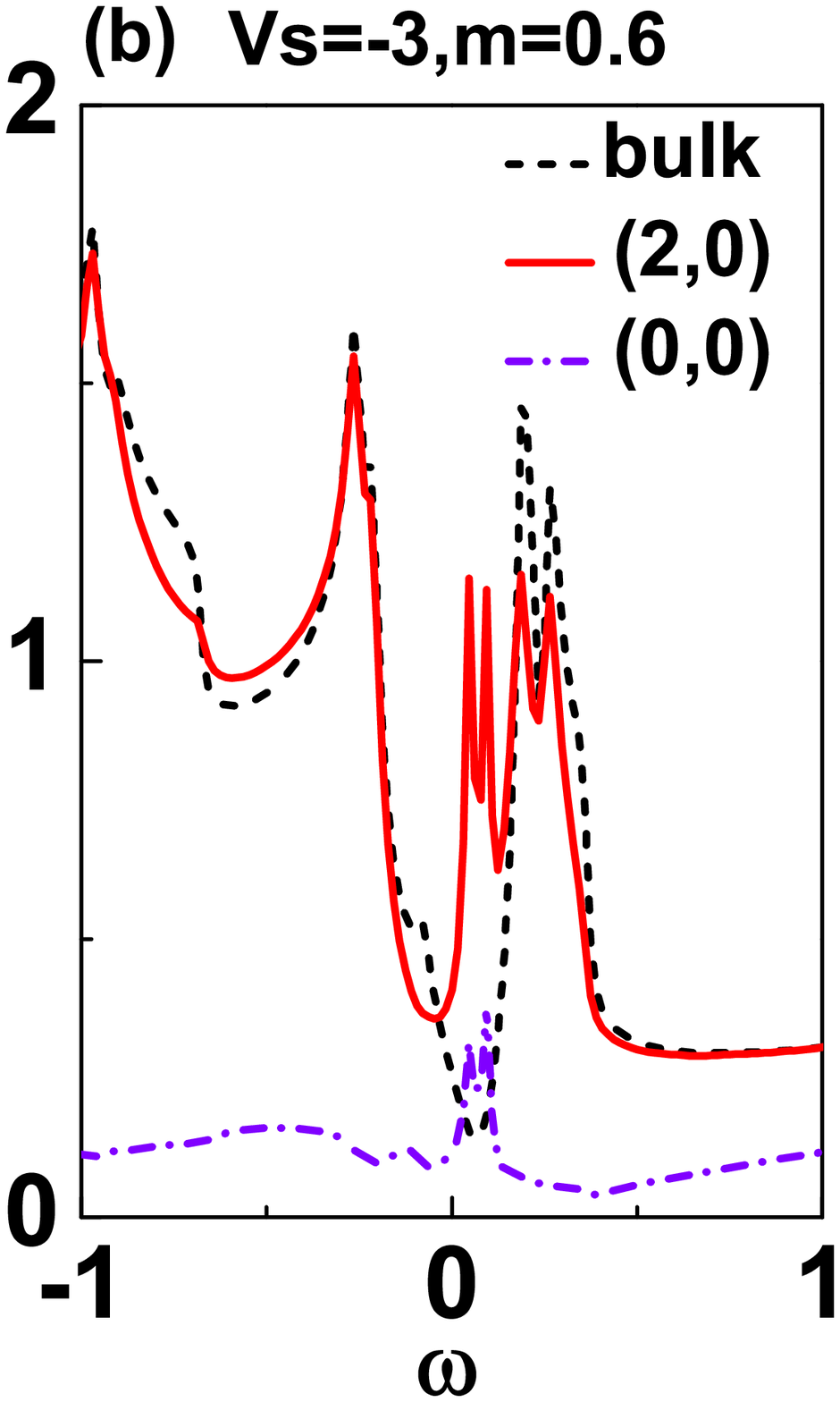}
\caption{(color online)Panel(a) shows LDOS at the different sites
for $Vs=3$ with $m=0.6$. Black dashed line represents LDOS in bulk,
the red solid line represents the site $(2,0)$, and the violet dash
dot line represents impurity site. Panel(b) is similar to (a) with
$Vs=-3$. } \label{fig13}
\end{figure}

\begin{figure}
      \includegraphics[width=1.68in]{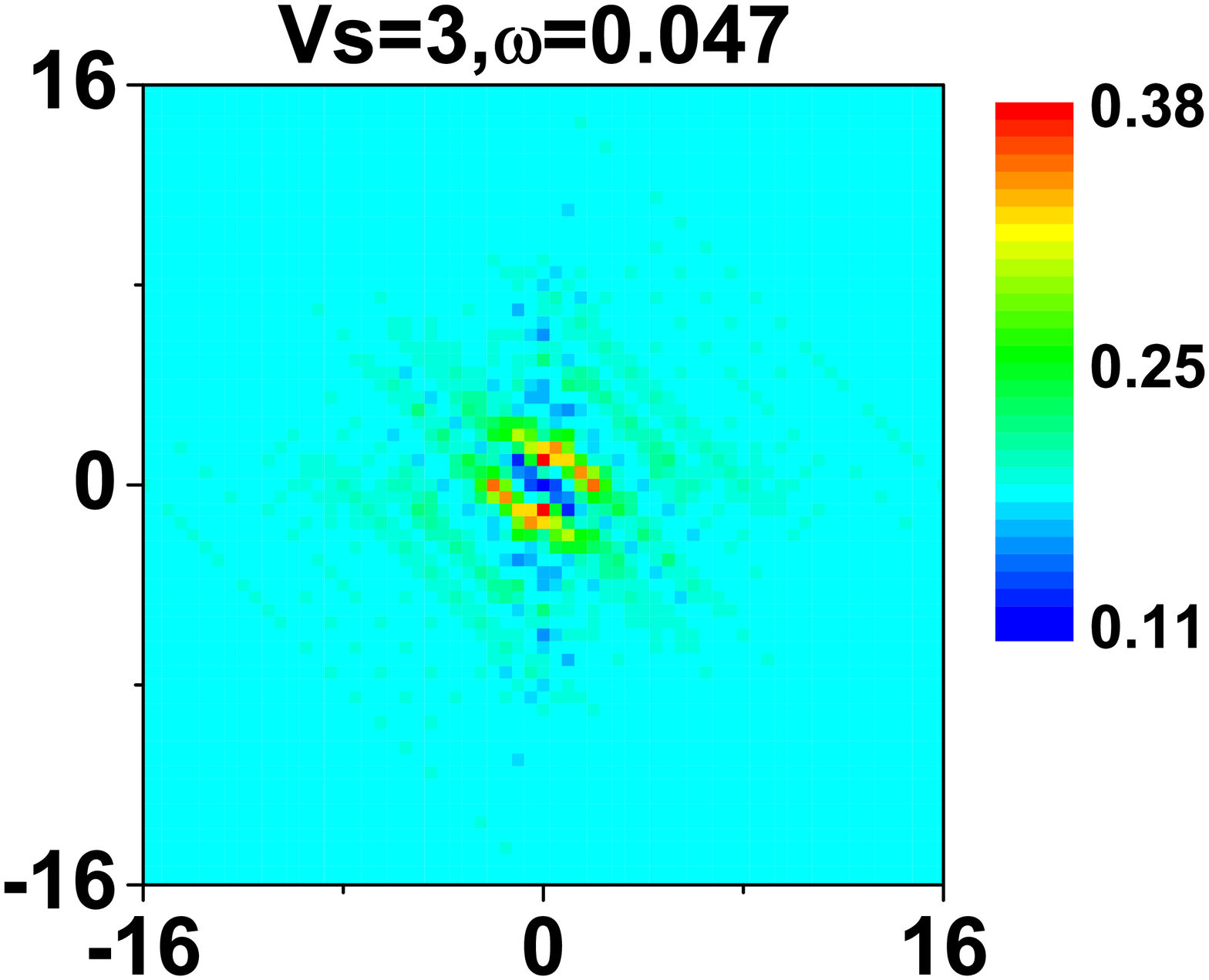}
      \includegraphics[width=1.68in]{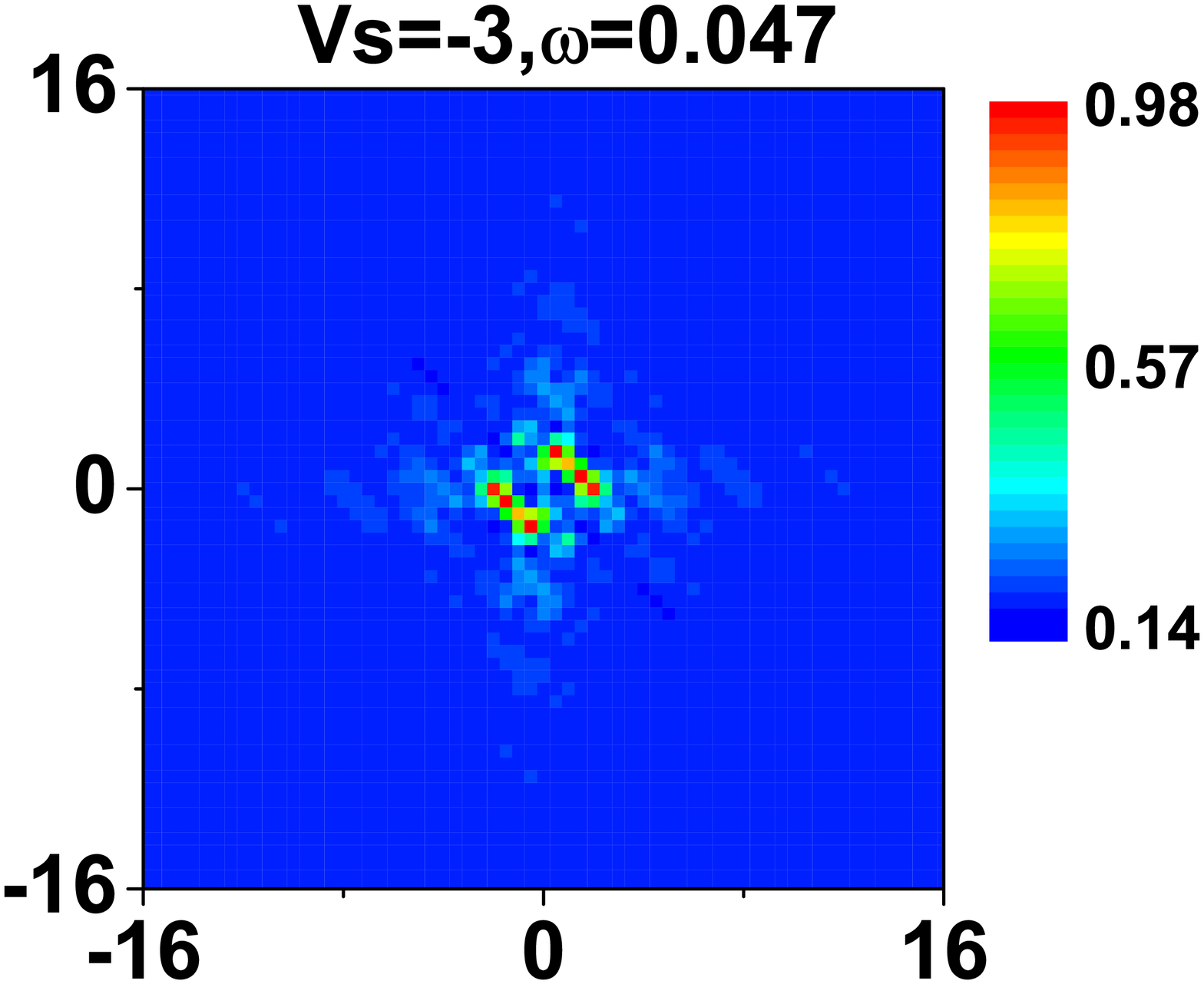}
      \includegraphics[width=1.68in]{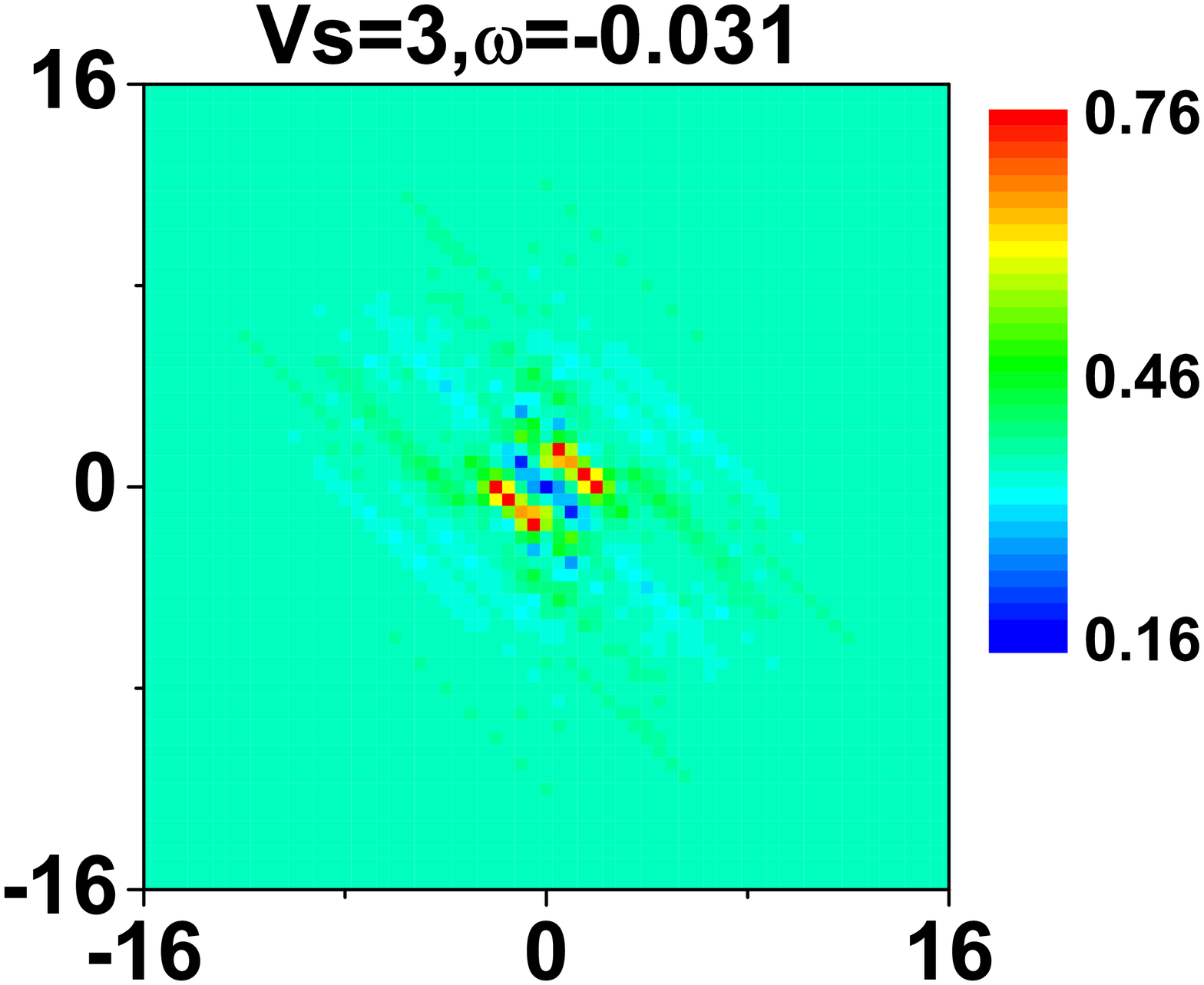}
      \includegraphics[width=1.68in]{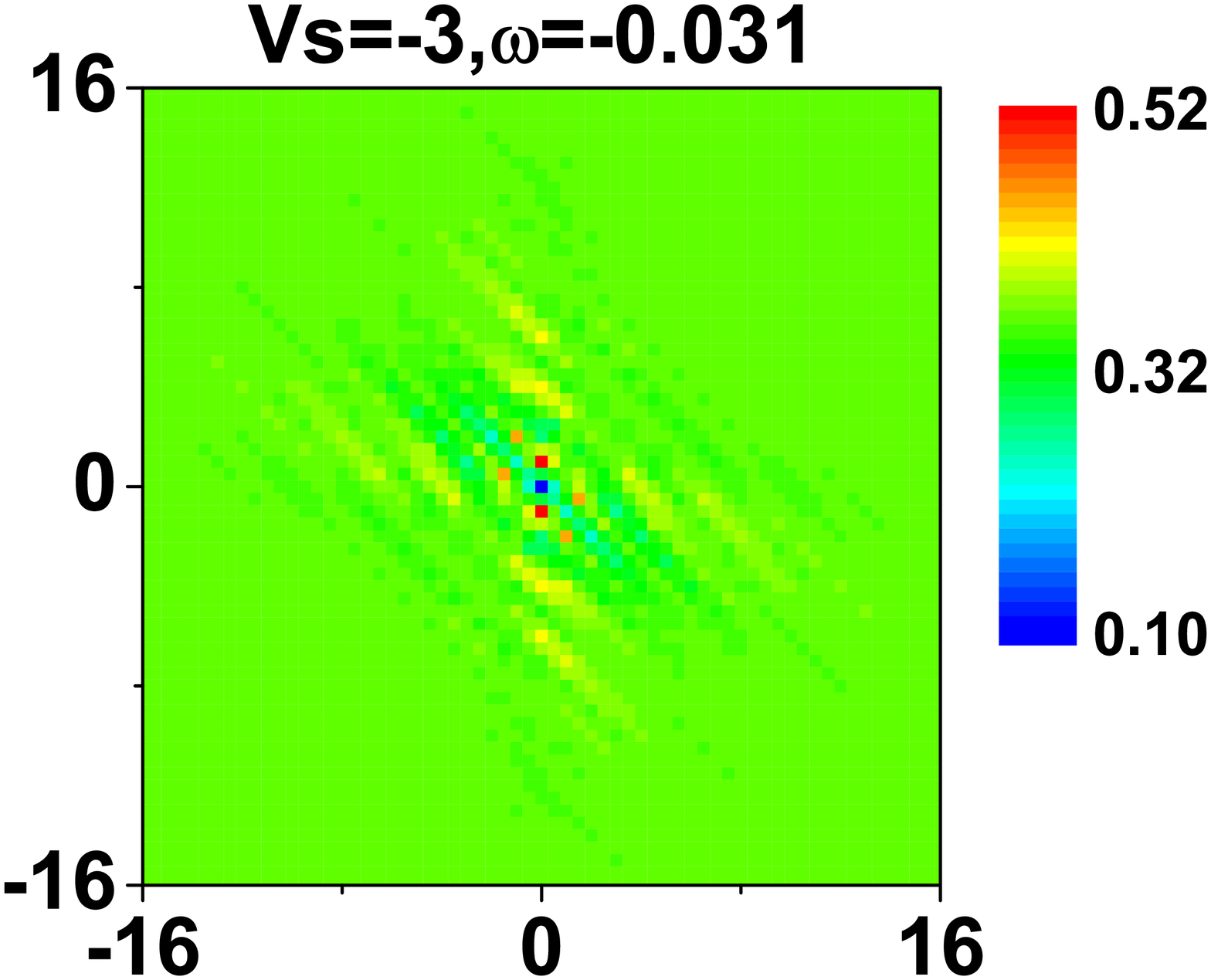}
\caption{(color online)Similar to Fig.\ref{fig11}, but for SP
$Vs=\pm 3$.} \label{fig14}
\end{figure}

\begin{figure}
      \includegraphics[width=1.55in]{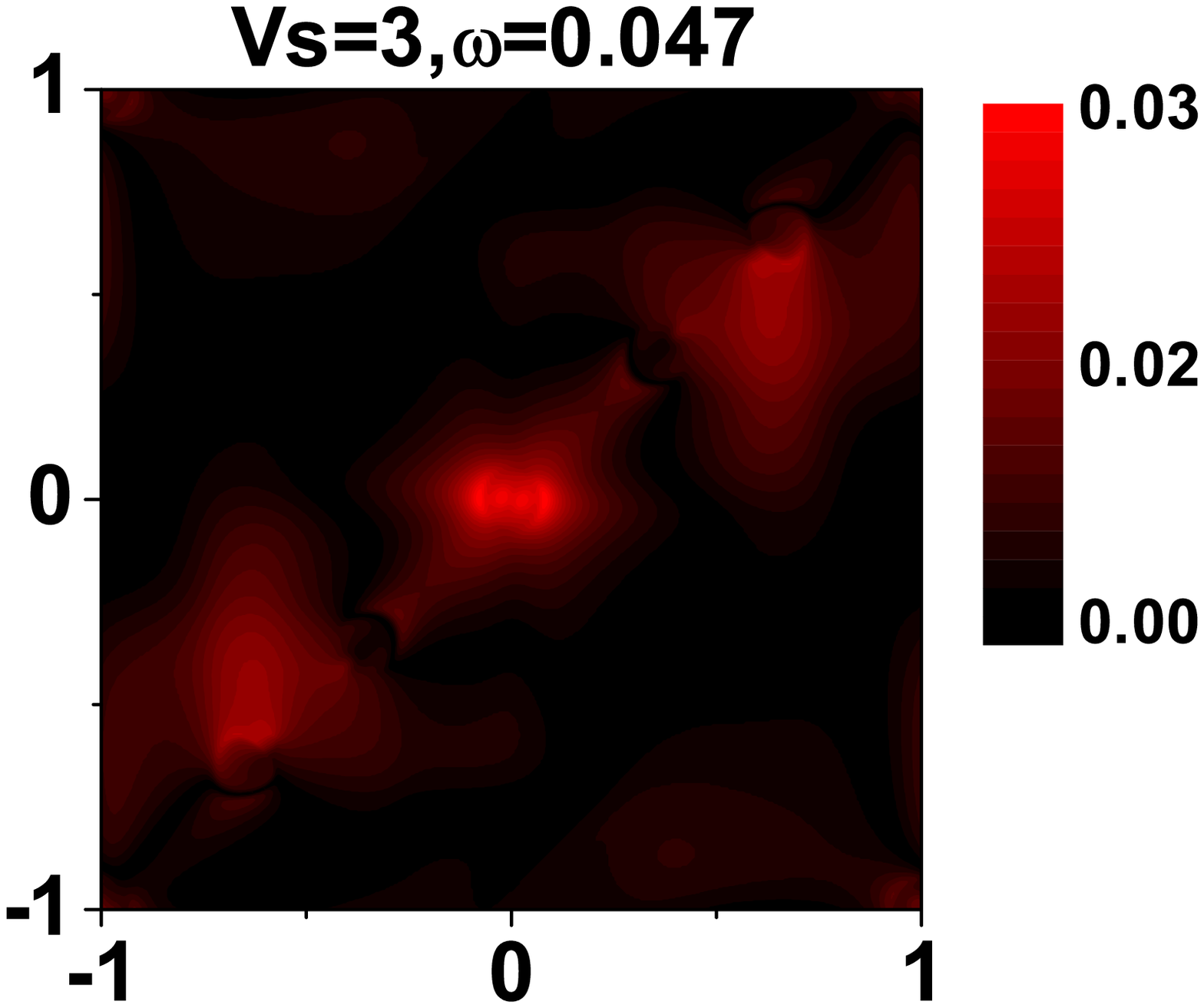}
      \includegraphics[width=1.68in]{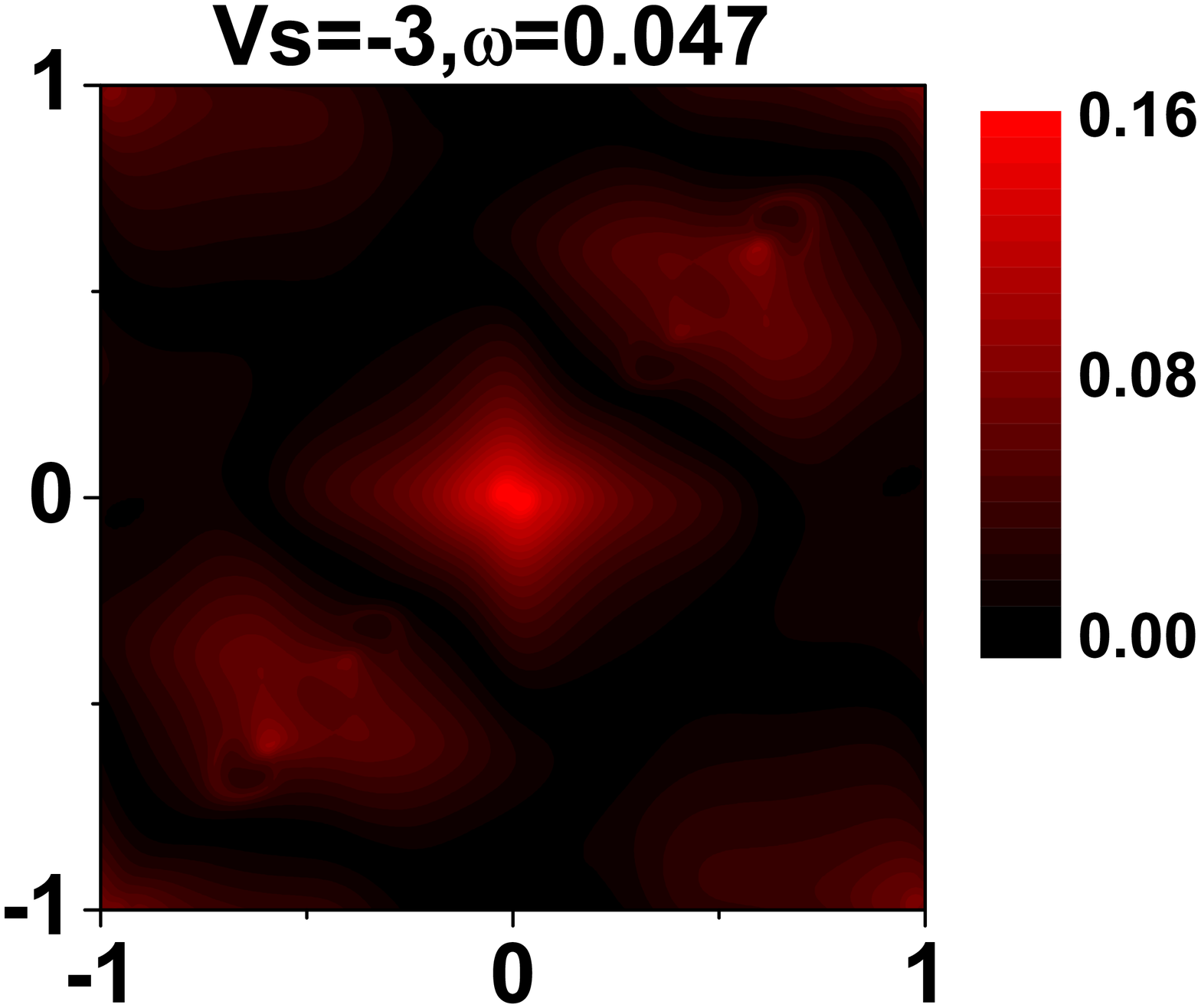}
      \includegraphics[width=1.55in]{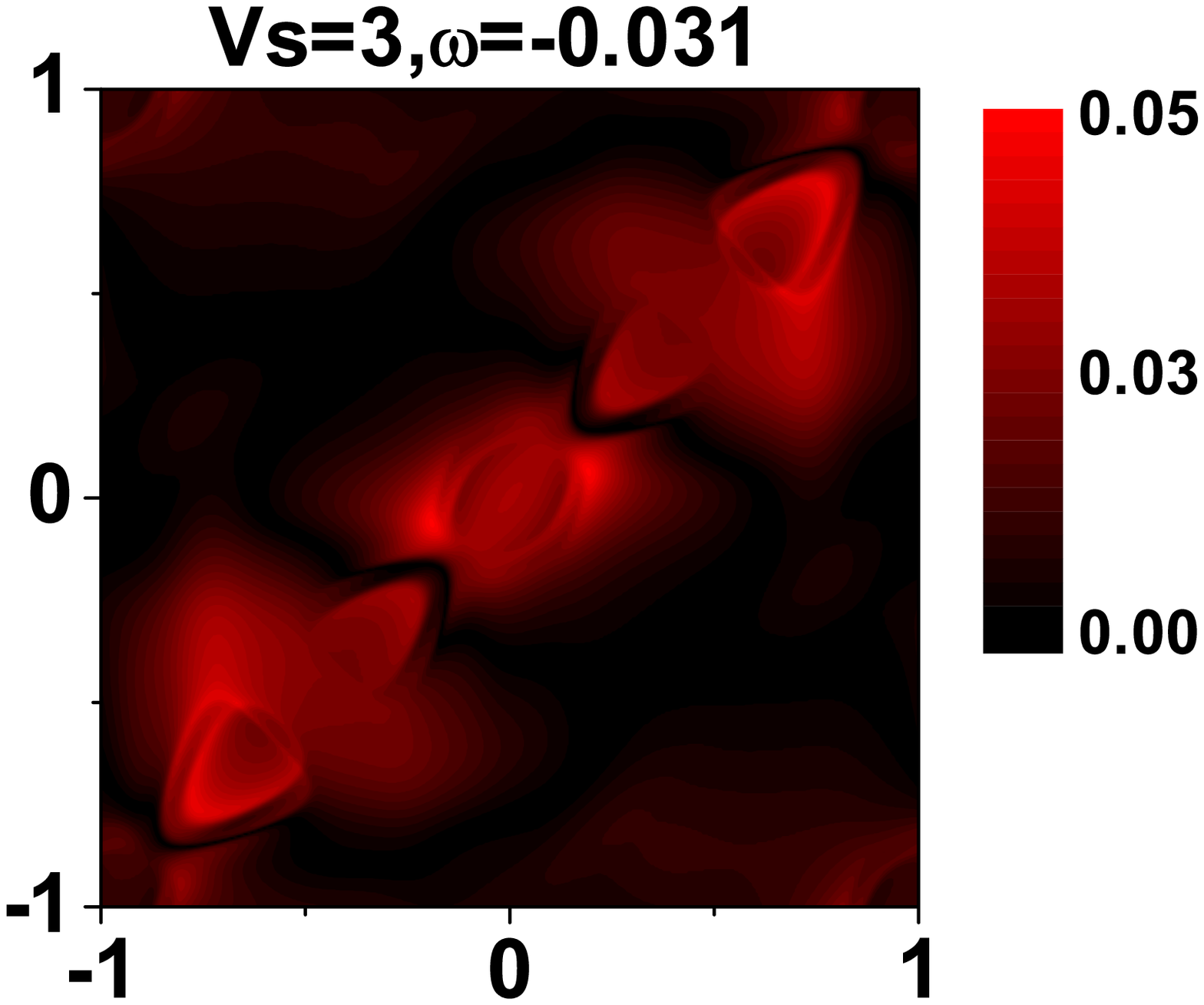}
      \includegraphics[width=1.68in]{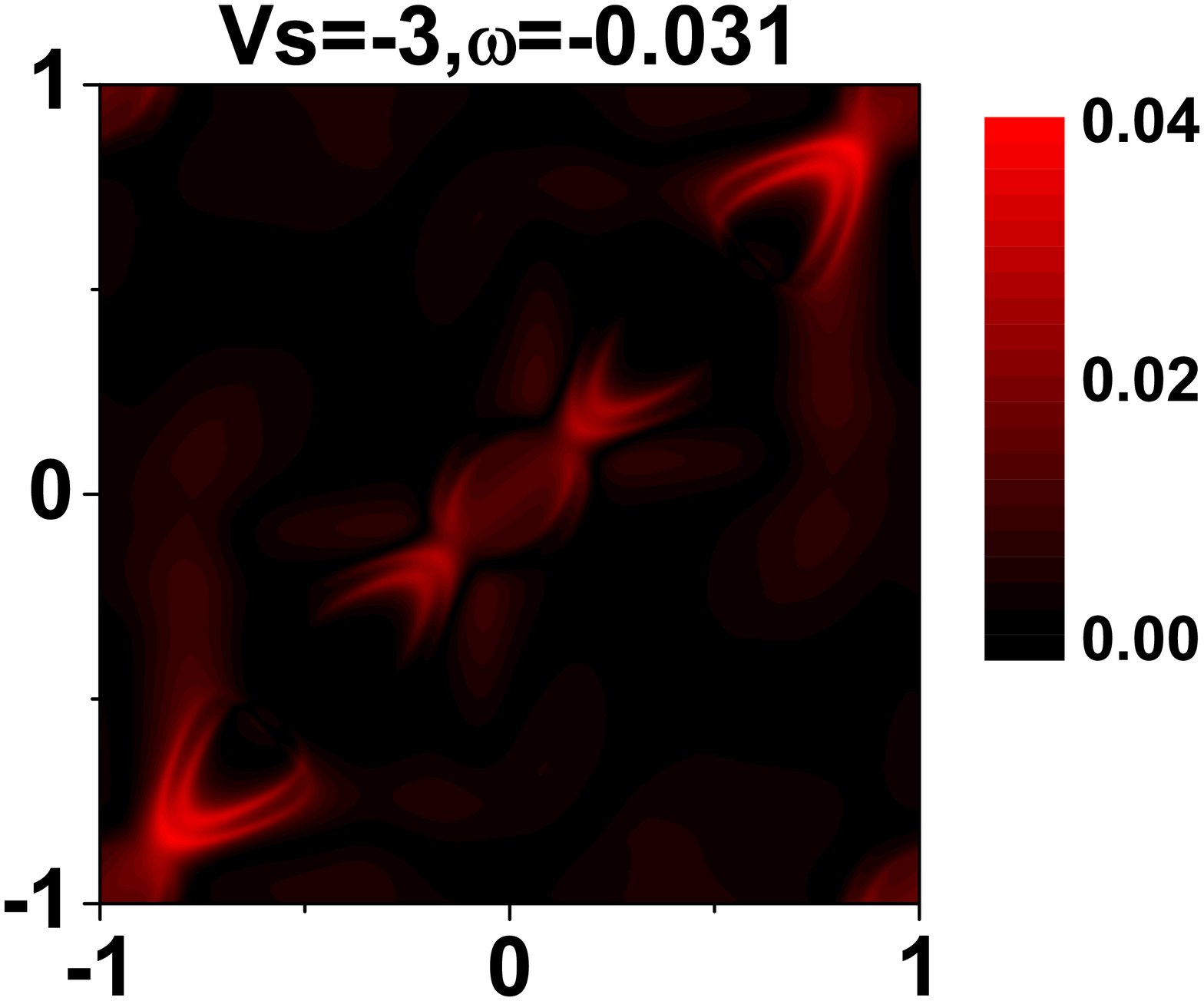}
\caption{(color online)Similar to Fig.\ref{fig12}, but for SP
$Vs=\pm 3$.  } \label{fig15}
\end{figure}

For strong SP $Vs=\pm3$, the LDOS at some sites in the vicinity of
the impurity is strongly affected. Pronounced in-gap resonance peaks
appear as shown in Fig.~\ref{fig13}. For $Vs=3$, one in-gap peak is
located at the negative energy $\omega=-0.031$. While there exist
two very close resonance peaks for $Vs=-3$ at the positive energies
$\omega=0.047$ and $0.094$, respectively. Those in-gap peaks reflect
the formation of bound states induced by QPI. We show the image map
of the LDOS in real space at selected energies $\omega=0.047$ and
$-0.031$ for $Vs=\pm3$ in Fig.~\ref{fig14}. We can see that, the 1D
stripe modulations of the LDOS are remarkable in all cases. The
sites with in-gap resonance peaks are located along the lines
$y=-x\pm2$. Fig.~\ref{fig15} shows $\rho_q(\omega)$ for
$m=0.6,Vs=\pm 3$. As can be seen, at all selected energies the QPI
wave vectors are along the diagonal direction, though they form
different patterns. For $Vs=3,\omega=-0.031$, the width of the
stripe-like pattern is apparently extended since the underlying band
structure plays a important role at energies away from zero. For
$Vs=-3,\omega=-0.031$, the dominate scattering $q$ is still about
$(\pm0.7\pi,\pm0.7\pi)$ thus the distance between two stripes in
real space is about $2a$.

For unitary case, LDOS is identity for positive and negative SP.
Our calculations show that at low energy
$\omega=0.016$, the spatial modulation of the LDOS still has stripe
pattern. However, at higher energy $\omega=0.187$,  beside the 1D
high-intensity stripes there are two circles around the impurity.
The dispersion pattern evolves with energy, at high energy
$\omega=0.187$ the pattern changes dramatically and has 2D
characteristics. Therefore, the dispersion relation is complex in
the multi-band system, and can not be fitted by a simple function as
in the cuprates~\cite{disper}.

\section{summary}
We have investigated by the T-matrix method the modulation of LDOS
and FC-LDOS in the SDW state of the iron-pnictides induced by QPI
for different impurity strength and well explained the two
experimental works[\onlinecite{chuang},\onlinecite{xiaodong}]. QPI
is sensitive to the value of the magnetic order which may vary from
one compound to another.

For small magnetic order, beside the high intensity small pockets
aligning along the diagonal direction, the zero-energy spectral
function exhibits high-intensity squares around the $\Gamma$ point,
therefore it is easy to form 2D QPI patterns. Our calculations show
that the 2D patterns of LDOS exist in real- and q-space for various
SPs. The exact pattern varies with the energy and in some cases QPI
induces ripple-like Friedel oscillations. It is consistent with what
have been observed in the $1111$ compound.~\cite{xiaodong}

For larger magnetic order, the main feature of the spatial
modulation of the LDOS is the 1D structure at the energies lower
than SDW gap. The QPI pattern in momentum space also supports the
formation of unidirectional nanostructures~\cite{chuang}. Negative
SP favors the inter-pocket scattering more than the repulsive one.
The LDOS on some sites in the vicinity of the impurity shows sharp
in-gap resonance peaks since the corresponding ungaped Fermi
surfaces are enlarged and scattering have more probability to induce
excitation at low energies. Our calculations show that remarkable 1D
stripe structure aligns along the FM direction in real space. The
reason is that for large-m system, zero-energy spectral function has
four isolated pockets along the AFM direction. In addition, the
topological analysis [\onlinecite{yingr}] of a two-band model and a
$5$-band model showed that the stable ungaped Fermi pockets are
along the AFM direction thus we expect the QPI obtained in those
models should have 1D stripe structure along the FM direction in
real space, similar to our results.

Our model has $C_4$ symmetry around As iron, both the impurity and
SDW could reduce the symmetry to $C_2$. The ungaped Fermi pockets as
well as the underlying band structure have contribution to the QPI
at bias energies away from zero. We obtain the 1D and 2D QPI
patterns observed by experiments based on one phenomenological
model, the microscopic origin of 1D and 2D QPI patterns is the shape
of spectral function at low energies.

\section{acknowledgements}

The authors would like to thank S. H. Pan, Ang Li and Jian Li for
useful discussions. This work was supported by the Texas Center for
Superconductivity at the University of Houston and by the Robert A.
Welch Foundation under Grant No. E-1146. Huang also acknowledges the
support of Shanghai Leading Academic Discipline Project S30105  and
Shanghai Education Development Project.

\end{document}